%% file: main.tex
\documentclass[conference,compsoc]{IEEEtran}
\usepackage{tikz}
\usepackage{amsmath}

\usepackage{filecontents}

\usepackage{balance}
\usepackage{algorithmic}
\usepackage{amsmath}
\usepackage{graphicx}
\usepackage{textcomp}
\usepackage{xcolor}
\usepackage[ruled,vlined,linesnumbered]{algorithm2e}
\usepackage{listings}
\usepackage{mdframed}
\usepackage{framed}
\usepackage{multicol}
\usepackage{multirow}
\usepackage{adjustbox}
\usepackage{enumitem}
\usepackage{colortbl}
\usepackage{threeparttable}
\usepackage{booktabs} % 可选，用于更好的表格线

\usepackage{tcolorbox}
\tcbuselibrary{most} % 加载一些常用的样式库

\newtcolorbox{mybox}[1][]{
    enhanced,
    frame hidden,
    boxrule=0.5mm,
    colback=yellow!20, % 背景色
    colframe=blue,     % 边框颜色
    sharp corners,
    #1
}

\newmdenv[
  backgroundcolor=gray!10, % 背景颜色
  linecolor=black,         % 边框颜色
  linewidth=1pt,           % 边框宽度
  leftmargin=5pt,          % 左侧外边距
  rightmargin=5pt,         % 右侧外边距
]{conclusionbox} % 自定义环境名称

\newcommand{\tool}{\textsc{RandLuzz}}

\newcommand{\eg}{\emph{e.g.}}

\newcommand{\etal}{\emph{et al.}}

\newcommand{\rowcolorA}{\rowcolor[gray]{0.9}} % 浅灰色
\ifCLASSOPTIONcompsoc
  \usepackage[nocompress]{cite}
\else
  \usepackage{cite}
\fi
\ifCLASSINFOpdf
\else
\fi
\begin{document}
%
% paper title
% Titles are generally capitalized except for words such as a, an, and, as,
% at, but, by, for, in, nor, of, on, or, the, to and up, which are usually
% not capitalized unless they are the first or last word of the title.
% Linebreaks \\ can be used within to get better formatting as desired.
% Do not put math or special symbols in the title.

\title{\Large \bf Large Language Model Assisted Directed Fuzzing:\\Understanding First, Then Fuzzing}
% \title{Further Automate Directed Fuzzing via Large Language Model}
% \title{Is Current Fuzzing Really Automated? \\Automating Directed Fuzzing Process via Large Language Models}
\title{Fuzzing: Randomness? Reasoning! \\Efficient Directed Fuzzing via Large Language Models}

% author names and affiliations
% use a multiple column layout for up to three different
% affiliations
\author{\IEEEauthorblockN{Xiaotao Feng}
\IEEEauthorblockA{360 Security Technology Inc.\\
Beijing, China\\
escofeng@gmail.com}
\and
\IEEEauthorblockN{Xiaogang Zhu}
\IEEEauthorblockA{School of Computer and Mathematical Sciences\\ The University of Adelaide\\
Adelaide, SA, Australia\\
xiaogang.zhu@adelaide.edu.au}
\and
\IEEEauthorblockN{Kun Hu}
\IEEEauthorblockA{School of Science\\ Edith Cowan University\\
Joondalup, WA, Australia\\
k.hu@ecu.edu.au}
\and
\IEEEauthorblockN{Jincheng Wang}
\IEEEauthorblockA{360 Security Technology Inc.\\
Beijing, China\\
wangjincheng1@360.cn}
\and
\IEEEauthorblockN{Yingjie Cao}
\IEEEauthorblockA{360 Security Technology Inc.\\
Beijing, China\\
yingjiecao@protonmail.com}
\and
\IEEEauthorblockN{Guang Gong}
\IEEEauthorblockA{360 Security Technology Inc.\\
Beijing, China\\
yingjiecao@protonmail.com}
\and
\IEEEauthorblockN{Jianfeng Pan}
\IEEEauthorblockA{360 Security Technology Inc.\\
Beijing, China\\
panjianfeng@360.cn}
}

\maketitle

% As a general rule, do not put math, special symbols or citations
% in the abstract
\begin{abstract}
Fuzzing is highly effective in detecting bugs due to the key contribution of randomness. However, randomness significantly reduces the efficiency of fuzzing, causing it to cost days or weeks to expose bugs.
Even though directed fuzzing reduces randomness by guiding fuzzing towards target buggy locations, the dilemma of randomness still challenges directed fuzzers.
Two critical components, which are seeds and mutators, contain randomness and are closely tied to the conditions required for triggering bugs.
% Therefore, to address the challenge of randomness, we propose to use large language models (LLMs) to remove the randomness for seeds in reaching target locations and reduce the randomness for mutators in examining buggy execution states.
Therefore, to address the challenge of randomness, we propose to use large language models (LLMs) to remove the randomness in seeds and reduce the randomness in mutators.
% With the excellent capability of reasoning and code generation, LLMs can be utilized to generate reachable seeds that can reach target locations and construct bug-specific mutators that are designed for target bugs.
With their strong reasoning and code generation capabilities, LLMs can be used to generate reachable seeds that target pre-determined locations and to construct bug-specific mutators tailored for specific bugs.
% With their strong reasoning and code generation capabilities, LLMs can be used to generate reachable seeds and to construct bug-specific mutators.
We propose \tool{}, which integrates LLMs and directed fuzzing, to improve the quality of seeds and mutators, resulting in efficient bug exposure. 
\tool{} analyzes function call chain or functionality to guide LLMs in generating reachable seeds.
To construct bug-specific mutators, \tool{} uses LLMs to perform bug analysis, obtaining information such as bug causes and mutation suggestions, which further help generate code that performs bug-specific mutations. 
% %%
% Even though directed fuzzing reduces randomness by guiding fuzzing towards target buggy locations, the randomness caused by seeds and mutation operators still significantly impact the efficiency.
% Therefore, we propose to use large language models (LLMs) to remove the randomness in reaching target locations and reduce the randomness in examining buggy execution states for directed fuzzing.
% With the excellent capability of reasoning and code generation, LLMs can be utilized to generate seeds that can reach target locations and construct mutators that are designed for specific bugs.
% We propose \tool{}, which integrates traditional static analysis of programs and LLMs, to generate inputs that are more likely to trigger bugs.
% Static analysis is utilized to extract required information for LLMs so that LLMs can focus on small and detailed tasks. This will improve the the quality of LLM-generated inputs, which can trigger bugs more efficiently.
%%%%%
We evaluate \tool{} by comparing it with four state-of-the-art directed fuzzers, AFLGo, Beacon, WindRanger, and SelectFuzz. 
With \tool-generated seeds, the fuzzers achieve an average speedup ranging from 2.1$\times$ to 4.8$\times$ compared to using widely-used initial seeds. Additionally, when evaluated on individual bugs, \tool{} achieves up to a 2.7$\times$ speedup compared to the second-fastest exposure.
On 8 bugs, \tool{} can even expose them within 60 seconds.

\end{abstract}

% no keywords

\input{body/intro}

\input{body/motivation}

\input{body/method}

\input{body/eval}

\input{body/discussion}
\input{body/relate}

\input{body/conclusion}

% For peer review papers, you can put extra information on the cover
% page as needed:
% \ifCLASSOPTIONpeerreview
% \begin{center} \bfseries EDICS Category: 3-BBND \end{center}
% \fi
%
% For peerreview papers, this IEEEtran command inserts a page break and
% creates the second title. It will be ignored for other modes.
\IEEEpeerreviewmaketitle

% % use section* for acknowledgment
% \ifCLASSOPTIONcompsoc
%   % The Computer Society usually uses the plural form
%   \section*{Acknowledgments}
% \else
%   % regular IEEE prefers the singular form
%   \section*{Acknowledgment}
% \fi

% The authors would like to thank...

%-------------------------------------------------------------------------------
\balance
\bibliographystyle{plain}
\bibliography{reference}

% that's all folks
\end{document}

%% file: body/intro.tex
\section{Introduction} \label{sec:Intro}

% \todo{(significantly) reduce randomness not replace; fuzzing still needs some randomness}
% \todo{\begin{itemize}
%     \item introduce the randomness in reachable seeds; why is it no randomness?
%     \item randomness in initial seeds: randomly collecting initial seeds. w cannot predict if initial seeds can reach target locations
% \end{itemize}}

Fuzzing has been proven highly effective at detecting bugs in real-world applications, with the discovery of numerous bugs across a wide range of software systems~\cite{zhu2022fuzzing, zhu2021regression, feng2021snipuzz}.
While existing fuzzers have introduced various strategies for fuzzing to trigger bugs, random input generation remains central to their effectiveness.
However, the other side of the coin is that the randomness often leads to inefficiency, causing fuzzers to run for days or even weeks to identify bugs located in deep code regions.

Typically, directed fuzzing~\cite{bohme2017directed} is more efficient than coverage-guided fuzzing~\cite{bohme2016coverage, zhu2021regression}; however, because directed fuzzing still relies on random input generation, it suffers from the same issue of inefficient bug discovery. Many solutions have been proposed to improve the efficiency of directed fuzzing~\cite{huang2024titan, xiang2024critical, huang2024everything}. 
Some directed fuzzers optimize the guidance towards target locations based on program properties, such as satisfying execution path constraints~\cite{lee2021constraint} or identifying deviation basic blocks~\cite{du2022windranger}. Another focus is generating inputs that are more likely to reach target locations via solutions such as predicting reachable inputs~\cite{zong2020fuzzguard, huang2024everything} and instrumenting related code only~\cite{luo2023selectfuzz}.
% However, all the existing solutions can only guide random input generation towards target locations.
Nonetheless, the existing directed fuzzing uses coarse-grained guidance to reduce the randomness, with a distance-based metric to guide the random input generation so that fuzzing can gradually reach target locations.

To improve the efficiency of triggering bugs, a straightforward yet challenging solution is to reduce or even remove the randomness when generating bug-triggering inputs.
To trigger a bug, an input needs to satisfy both reaching the buggy code location and examining specific execution states.
Thus, correspondingly, two critical components that significantly impact the efficacy of triggering bugs are the seeds and the mutation operators (mutators).
Seeds introduce randomness because the code regions they examine are unknown. They may be far away from or close to target locations.
If the seeds are already capable of reaching target locations, the efficiency of bug discovery can be greatly enhanced because it does not introduce randomness~\cite{klees2018evaluating}.
After reaching target locations, mutators are responsible of generating new inputs that examine diverse execution states for bugs. 
However, all existing works rely on pre-determined mutators, which are not designed for triggering specific bugs.
Such pre-determined mutators increase the randomness in fuzzing because most mutations cannot move states towards buggy ones~\cite{lyu2019mopt}.
The challenge is that, both generating \textit{reachable seeds}, which can reach target locations, and constructing \textit{bug-specific mutators}, which are designed for triggering specific bugs, require deep understanding of the target programs and bugs.

% Two critical steps that introduce randomness and significantly impact the efficacy of directed fuzzing are the pre-determined initial seeds and the mutation operators (mutators).
% If the initial seeds are already capable of reaching target locations, the efficiency of bug discovery can be greatly enhanced, underscoring the importance of seed selection~\cite{klees2018evaluating}.
% Mutators are responsible of generating new inputs, but all existing works rely on pre-determined mutators for mutation strategies, limiting the efficiency of triggering diverse bugs~\cite{lyu2019mopt}.
% \todo{These two steps also significantly contribute to bug triggering because initial seeds help reach target locations while mutators help satisfy execution states of bugs.}
% The challenge is that, both selecting reachable seeds, which can reach target locations, and constructing bug-specific mutators, which are designed for triggering specific bugs, require deep understanding of the program under test and the target bugs.

To address the challenge, in this paper, we propose to use large language models (LLMs) to remove the randomness in generating reachable seeds and reduce randomness in mutators.  
\textit{Our key insights are that i) LLMs~\cite{zhu2024when} possess strong capabilities in reasoning about bug information from both documentation and program code; and ii) the code generation capabilities of LLMs offer the potential to create adaptable and customizable mutators.} By leveraging these reasoning and code generation capabilities, LLMs can be used to generate reachable seeds and construct bug-specific mutators, enabling more efficient examination of buggy code regions. 
This will significantly reduce the impact caused by randomness in fuzzing.

To use LLMs for reducing randomness in fuzzing, the basic required information is the bug information, program usage, and function summary, which aim to understand the semantic meaning of bugs and the program logic of target projects.
To generate reachable seeds, because directly querying LLMs leads to exceeding token limits or misunderstanding, we query LLMs based on the Function Call Chain (FCC) that contains the paths from the entry point to the target vulnerable function.  
% By querying reachable seeds along the FCC, the generation process is efficient.
The FCC leads LLMs to generate reachable seeds step-by-step, which improves the effectiveness of generation process.
If we cannot obtain FCC, we will generate reachable seeds based on the functionalities of functions that are neighbors of the target function. 
If a seed can reach a neighbor function, we can use FCC to guide the generation of reachable seeds.
To generate bug-specific mutators, LLM is first utilized to analyze the cause for the target bug. The result is further utilized to generate mutation suggestions. Code is then generated for mutators based on the mutation suggestions via LLMs.

We develop \tool{}, which stands for \underline{Rand}om-\underline{L}ess F\underline{uzz}er, to demonstrate the performance of our method. 
We compare our \tool{} with four state-of-the-art fuzzers, AFLGo~\cite{bohme2017directed}, Beacon~\cite{huang2022beacon}, WindRanger~\cite{du2022windranger}, and SelectFuzz~\cite{luo2023selectfuzz}.
% The results show that \tool-generated seeds can significantly improve the efficiency of bug discovery, with a speedup ranging from 2.1$\times$ to 4.8$\times$ on average.
The results show that \tool-generated seeds significantly enhance bug discovery efficiency, achieving an average speedup ranging from 2.1$\times$ to 4.8$\times$.
% Moreover, \tool{} can expose 8 bugs within 60 seconds, leading the bug discovery efficiency.
Additionally, \tool{} can expose 8 bugs within 60 seconds, demonstrating leading efficiency in bug discovery.

Our contributions are as follows.
\begin{itemize}
    \item We analyze and demonstrate that randomness is the key factor that impacts the efficiency of directed fuzzing. 
    \item We introduce LLMs to remove randomness in generating reachable seeds and reduce randomness in generating bug-specific mutators.
    \item We develop \tool{}\footnote{Our code will be publicly available upon acceptance.} incorporating LLMs to demonstrate the effectiveness and efficiency of our idea, showing the feasibility of using LLMs in guiding directed fuzzing.
    \item Our evaluation demonstrates the effectiveness and efficiency of reachable seeds and bug-specific mutators.  
\end{itemize}

%% file: body/motivation.tex
\section{Motivation}\label{sec:motivation}

We identify two key components, which are seeds and mutators, that introduce randomness, reducing efficiency in directed fuzzing. 
In this section, we analyze such randomness based on the bug CVE-2020-13790.

% In real-world testing of various projects, we found that the effectiveness of directed fuzzing is influenced by the tester's understanding of the program being tested.
% This tester`s understanding is reflected in the selection (or construction) of the initial seeds for fuzzing and the design of mutation strategies.

% In this section, we will present how the selection of initial seeds and the design of mutation strategies influence directed fuzzing through a step-by-step demonstration of a directed fuzz testing based on CVE-2020-13790 project.
% For this fuzz testing, we have obtained the source code and compilation commands of the vulnerable version of the project.
% Our next step is to define the inputs and strategies relevant to the directed fuzz testing.

\subsection{Randomness in Seeds}

\begin{figure}[!t]
    \centering
    \includegraphics[width=0.85\linewidth]{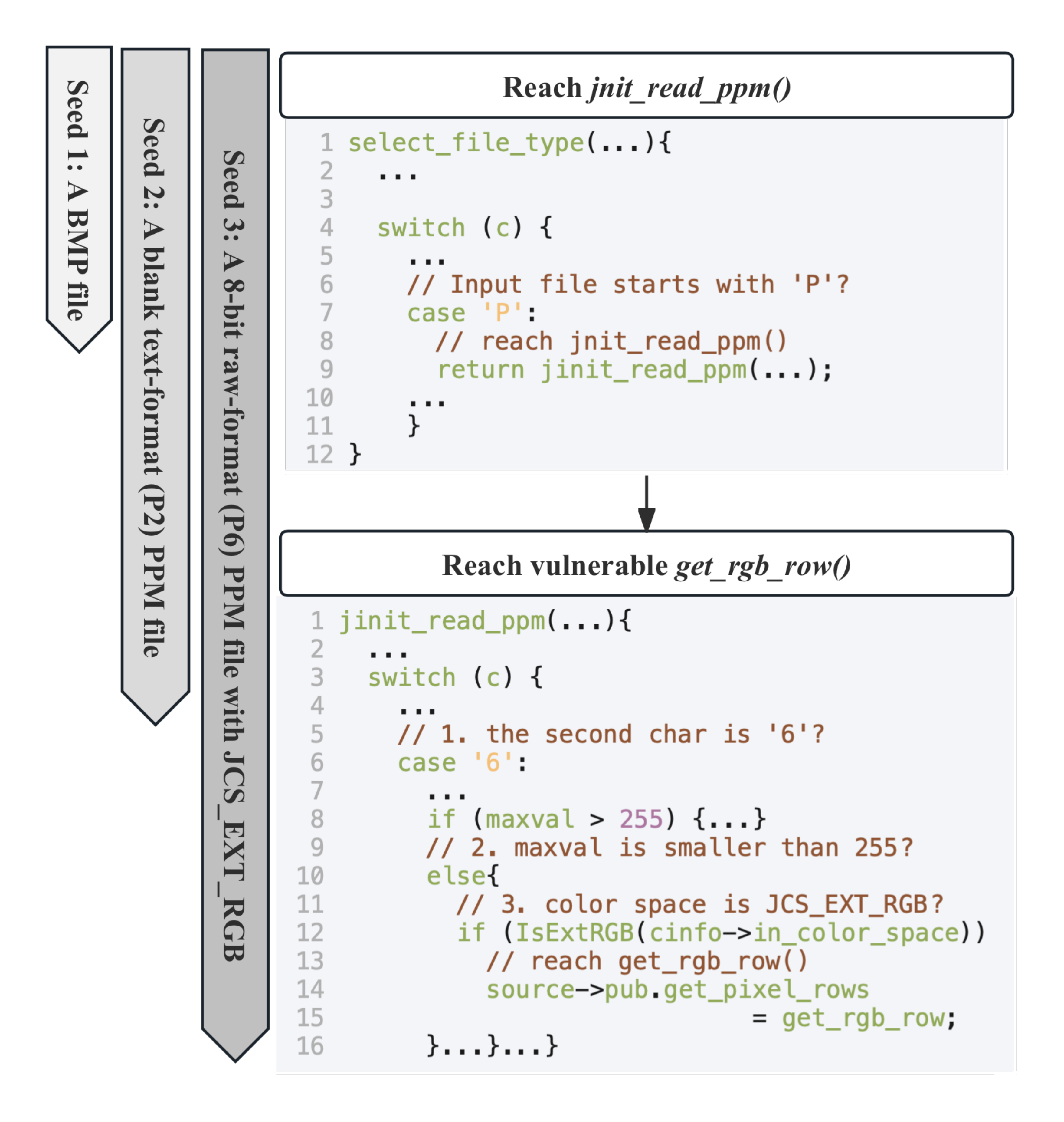}
    % \vspace{-7mm}
    \caption{Randomness in initial seeds for \texttt{cjpeg}. Different seeds explore different input space, resulting in differences in time to bugs.}
    \label{fig:motivation_eg}
\end{figure}

Figure~\ref{fig:motivation_eg} shows that three seeds with different file formats can reach different code regions in the program \texttt{cjpeg}.
To reach the target vulnerable function \texttt{get\_rgb\_row()}, a seed has to satisfy multiple path constraints.
In Figure~\ref{fig:motivation_eg}, the \textit{Seed 1} is a blank BMP image that can only examine the function \texttt{select\_file\_type()}. 
To reach the function \texttt{jinit\_read\_ppm()}, the \textit{Seed 2} satisfies a different file format PPM, whose input file starts with the characters \textit{P2}.
When trying to reach the vulnerable function \texttt{get\_rgb\_row()}, a more complex file format has to be satisfied.
Thus, the carefully crafted \textit{Seed 3}, which can reach \texttt{get\_rgb\_row()}, is an 8-bit raw-format (P6) PPM file with the \texttt{JCS\_EXT\_RGB} color space.

The randomness in the seeds lies in the fact that they may have different search space to explore.
For example, the \textit{Seed 1} has to explore the input space to satisfy all path constraints including the bytes specifying the file format, file size, and color space.
However, for \textit{Seed 3}, since it already satisfies all the path constraints, it can focus on exploring the buggy execution states.
% This will significantly reduce the randomness in moving towards buggy states.
To showcase the observation, we test the three seeds on the bug CVE-2020-13790 with the directed fuzzer AFLGo. AFLGo runs each seed four times and their timeouts are set to 24 hours. 
% If a seed cannot discover the bug within 24 hours, we will terminate the testing.
The results show that \textit{Seed 1} cannot expose the bug due to its large search space and randomness.
\textit{Seed 2} can only discover the bug once, with the testing time close to 24 hours (84,509 seconds).
However, \textit{Seed 3} can identify the bug twice, and it exposes the bug more quickly, with the testing time close to 8 hours (30,211
seconds) and 5 hours (17,828 seconds).

\begin{figure*}[!t]
    \centering
    \includegraphics[width=0.95\textwidth]{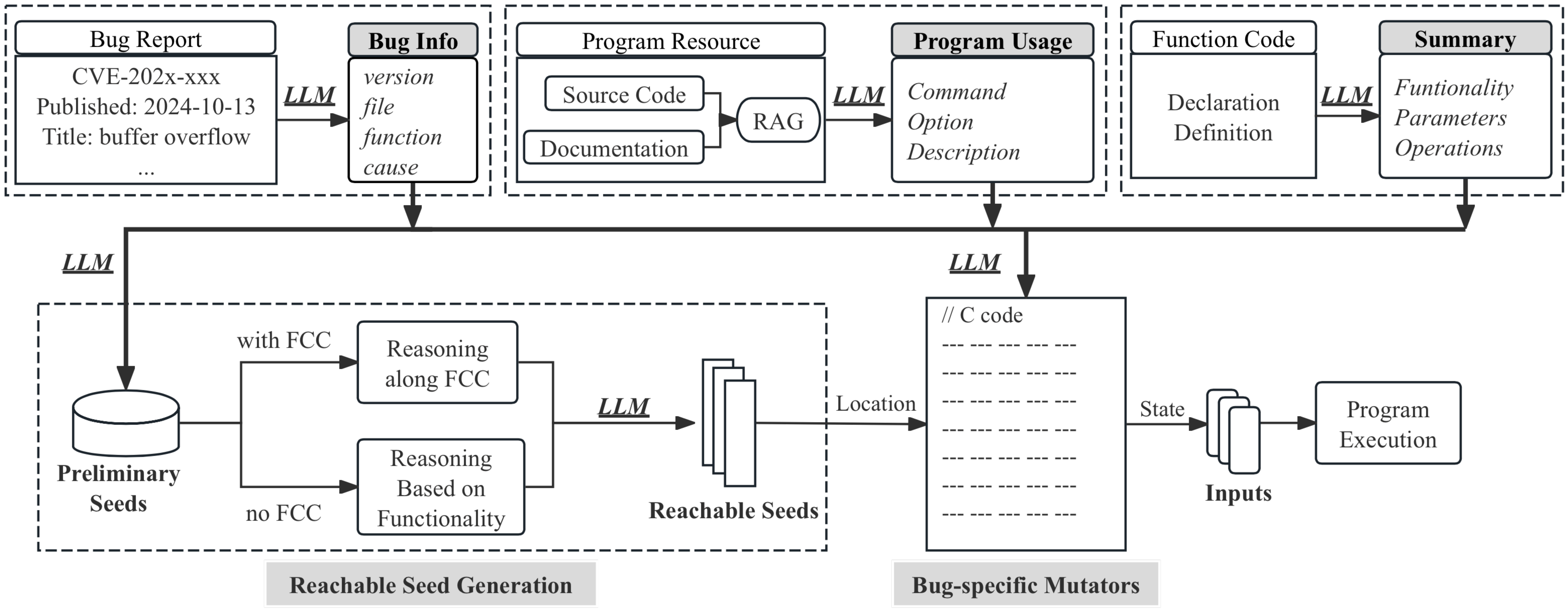}
    % \vspace{-7mm}
    \caption{Workflow of \tool{}. \textit{Reachable} indicates that the execution can reach the target vulnerable location. The use of LLM is to generate reachable seeds and generate specific mutators for target bugs. FCC is short for Function Call Chain.}
    \label{fig:workflow}
\end{figure*}

The challenge is that, effective seeds require expertise to craft, reflecting various levels of understanding about the target programs and bugs.
% For example, a tester who only knows the target program is \texttt{cjpeg}, with no knowledge of the specific vulnerability, might use an image in any format (such as BMP) for testing.
For instance, a tester who knows only that the target program is \texttt{cjpeg}, without any specific knowledge of the vulnerability, might use an image in any format (such as BMP) for testing.
In essence, the tester’s level of understanding about the target programs and bugs significantly impacts the effectiveness of directed fuzzing.
Therefore, in this paper, we use LLMs to automatically understand target programs and bugs, which further helps craft reachable seeds for fuzzing.

\subsection{Randomness in Mutators}

Randomness in mutators inherits from the concept of fuzzing, which \textit{randomly} generates numerous inputs to repeatedly test target programs.
This immediately leads to the dilemma that fuzzing needs certain degree of randomness to maintain its effectiveness while randomness is expected to be reduced for efficiency consideration.
Existing mutators satisfy the requirement of randomness because they follow the idea of building meaningful mutations based on small and basic operations.
For example, most fuzzers use the mutators from AFL, which designs basic operations such as \textit{bit flips} and \textit{byte insertion}. 
% Combination of such operations may generate any meaningful mutations, but with the drawback of high randomness.
The combination of such operations can produce meaningful mutations but comes with the drawback of high randomness.
As shown in Figure~\ref{fig:motivation_eg}, \textit{Seed 3} can already reach target locations, but the time to bug still experiences discrepancies among different trials.

To trigger a bug, the mutation is required to generate inputs that examine specific locations and exercise certain execution states.
Most existing papers intend to efficiently apply mutators in exploring diverse code locations~\cite{bohme2016coverage, zhu2020csi, zhang2024shapfuzz}.
However, even a seed can reach buggy code regions, it still requires the mutators to explore execution states. 
We observe that, \textit{efficient mutators for exploring code regions and execution states are likely to be different}.
For example, in Figure~\ref{fig:motivation_eg}, to reach target function, the mutators have to satisfy path constraints such as inputs starting with characters \textit{P6}.
Yet, to satisfy the buggy execution states, the mutators need to manipulate memories to overflow the boundary of a buffer.
Therefore, in this paper, we design two solutions to reach target locations and to explore buggy states, respectively.
To reduce the randomness in mutators, our design focuses on constructing bug-specific mutators.

%% file: body/method.tex
\section{Methodology} \label{sec:method}

Our \tool{} aims to minimize the impact of randomness in fuzzing by leveraging LLMs. 
To effectively use LLMs in directed fuzzing, our \tool{} consists of multiple phases in using LLMs, as shown in Figure~\ref{fig:workflow}. The first phase is to use LLMs in preparing necessary information, which is utilized in later reasoning stages.
Then, \tool{} generates reachable seeds via reasoning based on the necessary information, which removes the randomness in reaching target locations. 
Finally, \tool{} constructs bug-specific mutators to explore execution states in target vulnerable locations, which reduces randomness in generating bug-triggering inputs.
% \todo{describe template}

% including \textit{Bug Information}, \textit{Program Usage}, and \textit{Function Summary}.
% \textit{Bug Information} contains the target vulnerable location and other information useful for generating inputs.
% \textit{Program Usage} is used to obtain the program options that can reach target vulnerable locations. \textit{Function Summary} is performed when necessary, and provides essential information for generating efficient inputs, including the functionality, parameters and operations.
% Then, \tool{} generates reachable seeds based on the aforementioned necessary information. Finally, \tool{} constructs bug-specific mutators to explore execution states in target vulnerable locations.

\subsection{LLM Query Scheme}

\begin{figure*}
    \centering
    \includegraphics[width=0.8\linewidth]{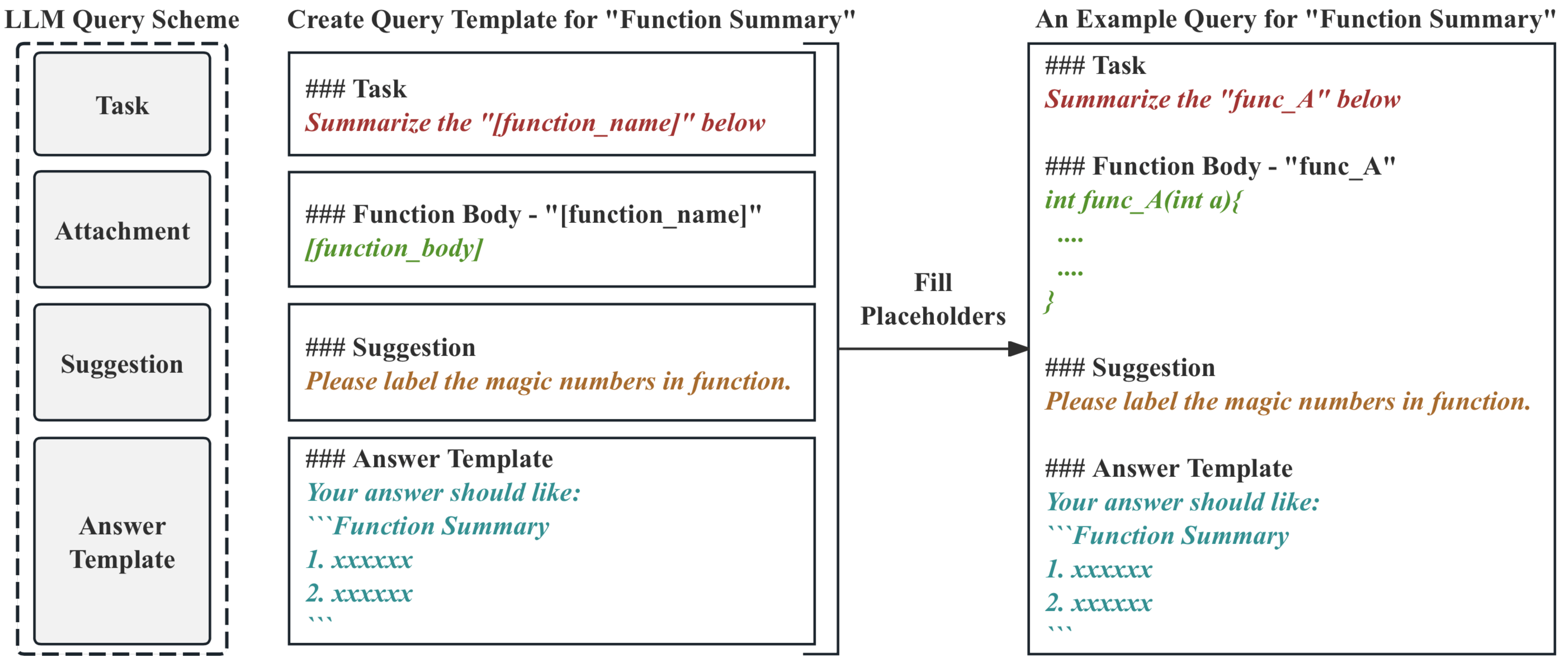}
    % \vspace{-7mm}
    \caption{LLM query scheme. Each task is based on this query scheme to create a query template, which further generates queries. }
    \label{fig:query_scheme}
\end{figure*}

Throughout the workflow, \tool{} frequently communicates with LLMs.
We devise an LLM query scheme, illustrated in Figure~\ref{fig:query_scheme}, which primarily consists of four parts:
\begin{itemize}[left=0pt]

    \item  \textbf{Task} describes the main objective of the query.

    \item  \textbf{Attachment} includes relevant context for the query, \eg, function body when querying function summary.
    
    \item  \textbf{Suggestion} provides LLMs with guidance relevant to the specific tasks and required answers.
    
    \item  \textbf{Answer Template} defines the required content and format of the LLM's response.

\end{itemize}

When faced with different tasks, \tool{} uses this query scheme to create task-specific templates, with the corresponding context. 
We design various query templates under the query scheme to handle a range of tasks.
An example query template is shown in Figure~\ref{fig:query_scheme}. When \tool{} needs to summarize a function in the program, it provides the scheme with the \textit{Task} \textit{``Summarize the function''}, the relevant context, as well as placeholders for detailed query. %, thereby generating the text for the query.
Notably, to reduce query time and the number of tokens used, communication between \tool{} and LLMs excludes any historical conversation records.
If a task requires results from a previous communication, \tool{} attaches these results in the \textit{Attachment} and reminds the LLMs of the presence in the \textit{Suggestion}.

\subsection{LLM-Assisted Necessary Information}

As shown in Figure~\ref{fig:workflow}, the necessary information includes \textit{Bug Information}, \textit{Program Usage}, and \textit{Function Summary}.
To perform directed fuzzing, the essential information includes the target code locations. Given a bug report, \tool{} automatically extracts bug related details (\textit{Bug Information}), such as target locations and other information useful for generating inputs. 
Next, the command options needed to reach the buggy locations in the program should be identified. 
\textit{Program Usage} is used to obtain the command options that can reach target vulnerable locations.
Finally, summaries of each function (\textit{Function Summary}) are necessary to understand the program, which can further assist in generating reachable seeds.
\textit{Function Summary} is performed when necessary and provides essential information, including the functionality, parameters and operations.

\subsubsection{Bug Information} \label{sec:bug_info}

Bug reports may have various formats to present bug related information. For example, a Common Vulnerabilities and Exposures (CVE)\footnote{https://cve.mitre.org/} report is a standardized document that provides detailed information about a specific security vulnerability or exposure, including its description, the affected software versions, and the potential impact.
Generally, the aim is to extract necessary information from bug reports, including the affected software versions, the location of the vulnerability (including the specific file and function), and a brief summary of the cause of the vulnerability, as shown in Figure~\ref{fig:workflow}. 
We leverage LLMs to extract such information because LLMs excel at summarizing and synthesizing natural language contents, enabling an approach to efficiently analyze CVE reports.

The query template for analyzing CVE reports is loaded to generate the appropriate query for communicating with the LLM. The \textit{Task} directly prompts the LLM to extract the required bug information from the CVE report, which is included as an \textit{Attachment}. The extracted bug information must be provided in a specified format, as outlined in the \textit{Answer Template}.
The bug information helps  pinpoint the location of bugs, serving as the target for subsequent optimization efforts. Additionally, details on the causes of bugs, combined with definitions of the functions where they reside, allow the LLM to analyze and identify the conditions under which the bugs can be triggered. This insight aids in developing mutators tailored to target bugs.

\subsubsection{Program Usage}\label{sec:program_use}

\begin{figure*}
    \centering
    \includegraphics[width=\linewidth]{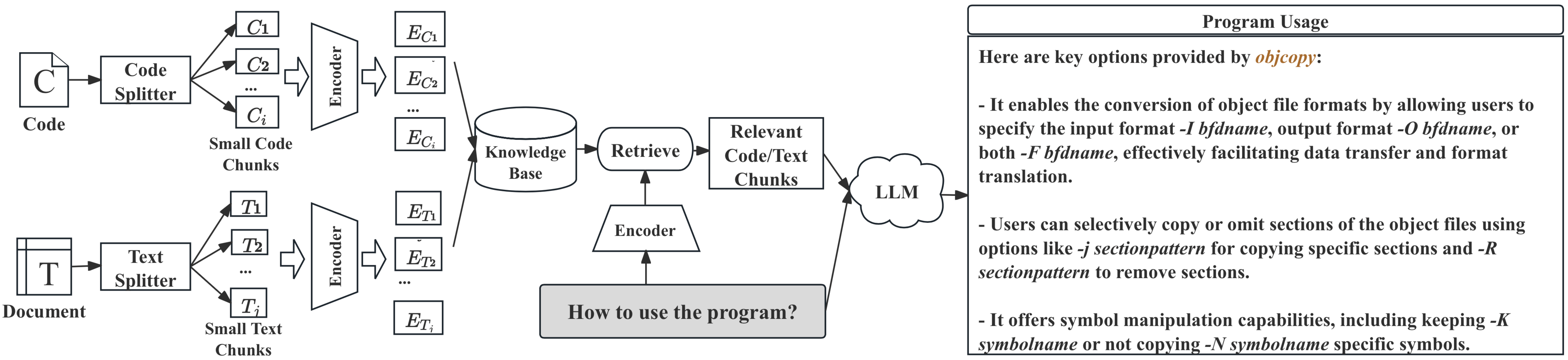}
    % \vspace{-7mm}
    \caption{Program usage and its options. \tool{} uses RAG to retrieve relevant and small chunks from documents and source code, and the small chunks are used as the context when querying LLMs.}
    \label{fig:usage_rag}
\end{figure*}

The program usage refers to the command lines required to run the target program, particularly the command options. Command options are often tailored for specific functionalities, indicating that different options exercise different code regions~\cite{song2020crfuzz}. 
For example, the program \texttt{objcopy} specifies the input format with the option \textit{-I} and the output format with the option \textit{-O}.
Some bugs, in fact, can only be exposed with specific options.
Extracting program usage by inputting the entire program as the prompt to LLMs is typically not feasible, as it  exceeds the token limit. Even if an LLM could handle an entire project, its output would likely be imprecise due to inherent challenges relevant to focus and coherence, memory and retention issues and increased computational cost associated with
the complexity of programs.
Therefore, we employ a Retrieval-Augmented Generation (RAG)-like strategy~\cite{lewis2020retrieval} to help LLMs gain in-depth understanding of a program, which in turn supports the generation of program usage. 
% \zhu{briefly introduce what is RAG \cite{}.}

As illustrated in Figure~\ref{fig:usage_rag}, RAG enhances LLMs by integrating with an external knowledge retrieval mechanism that includes a retriever, which sources pertinent information, and a generator, which synthesizes this information into coherent output. 
In our study, the process begins with bug information as a query, triggering the retriever to search a \textit{knowledge base} - the program project and its documentation repository - for relevant cues. 
Retrieval is based on the embedding similarity between the bug information and chunks of program code and documentation, divided into smaller parts.
Specifically, we use a sentence transformer - all-MiniLM-L6-v2~\cite{wang2020minilm} to map each chunk to a 384-dimensional dense vector embedding space, with Faiss employed for vector similarity search~\cite{douze2024faiss}.  
Empirically, we select the top 10 relevant chunks as the supplemental information. 
The retrieved supplemental information, together with the bug information, is then passed to the generator module, which in our case is the LLM. 
We query the LLM by setting a specific \textit{Task}, such as \textit{``Summarize the usage of all command options for this program''}, enabling it to generate a comprehensive response about program usage. Additionally, the query’s \textit{Answer Template} requires the LLM's response to include the command options mentioned in the chunk, along with their corresponding usage descriptions.
In later stages, the program usage information will guide the LLM in selecting appropriate command options, helping ensure that the target functions are reached.

\subsubsection{Function Summary}\label{sec:func_summary}

\begin{figure}
    \centering
    \includegraphics[width=\linewidth]{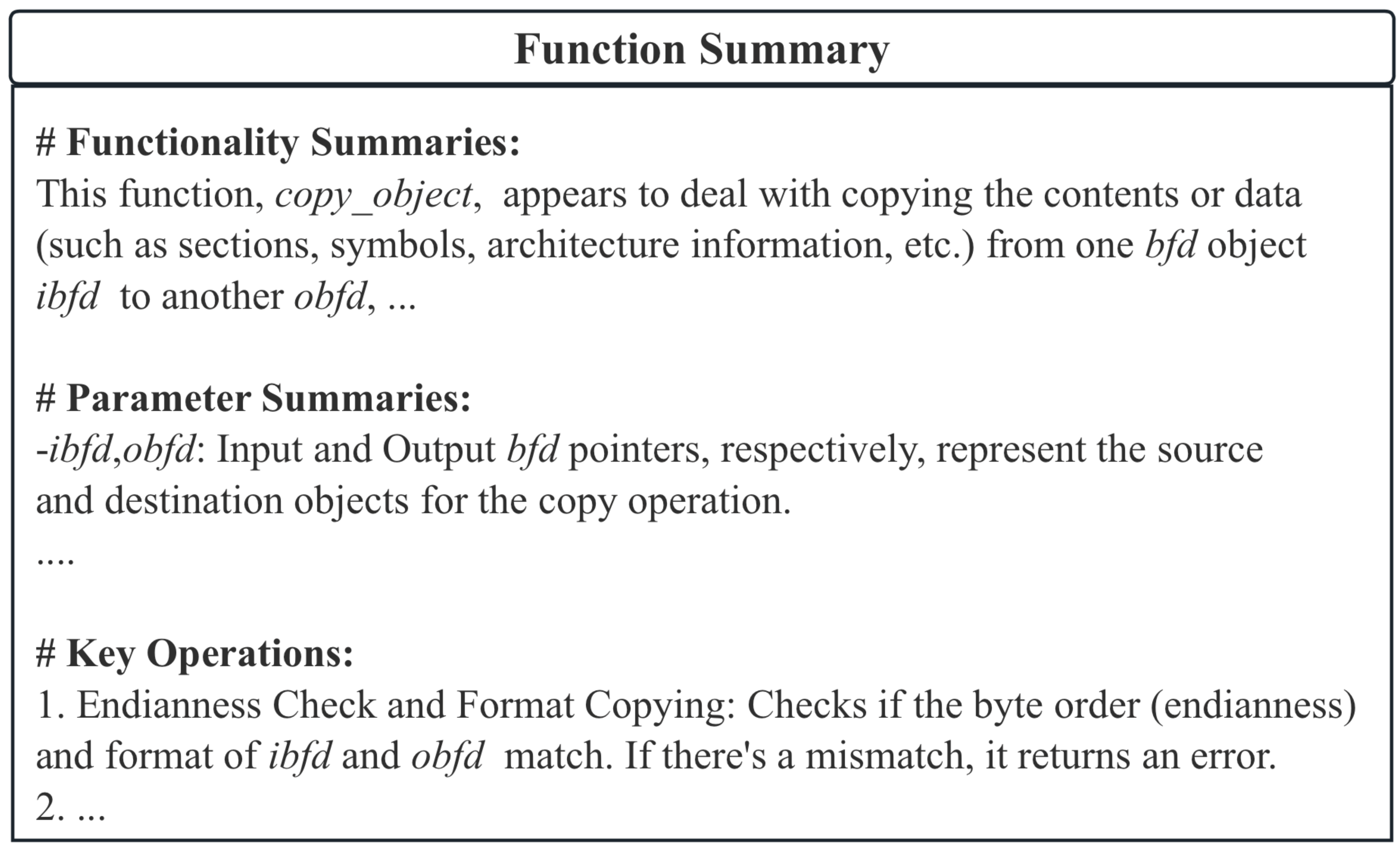}
    % \vspace{-5mm}
    \caption{An example of Function Summary. It includes the summaries of functionality, parameters, and key operations.}
    \label{fig:func_sum}
\end{figure}

The function summary report is essential for \tool{}, which helps infer relationships between functions and program command options. Additionally, key operations and variables within the function body can be leveraged for further seed optimization and vulnerability cause analysis.
\tool{} uses Clang's AST (Abstract Syntax Tree) component to analyze the project's source files. The main objective of this static analysis is to capture and record all functions within the project and map their call relationships, which are identified via simple control flow analysis, such as capturing keywords CALL\_EXPR or POINTER pointing to a FUNCTIONPROTO in the AST output. Additionally, if \tool{} requires more detailed information about a function, such as its specific functionality or variable values, it can query the LLM for an in-depth analysis.
With the function's declaration and definition extracted from static analysis included as the context in \textit{Attachment}, \tool{} queries the LLM by setting a \textit{Task} like \textit{``Summarize the function's purpose''}. The LLM's output will provide summaries of the function's functionality, parameters, and key operations as specified in the \textit{Answer Template}, as illustrated in Figure~\ref{fig:func_sum}.

\begin{figure*}
    \centering
    \includegraphics[width=\linewidth]{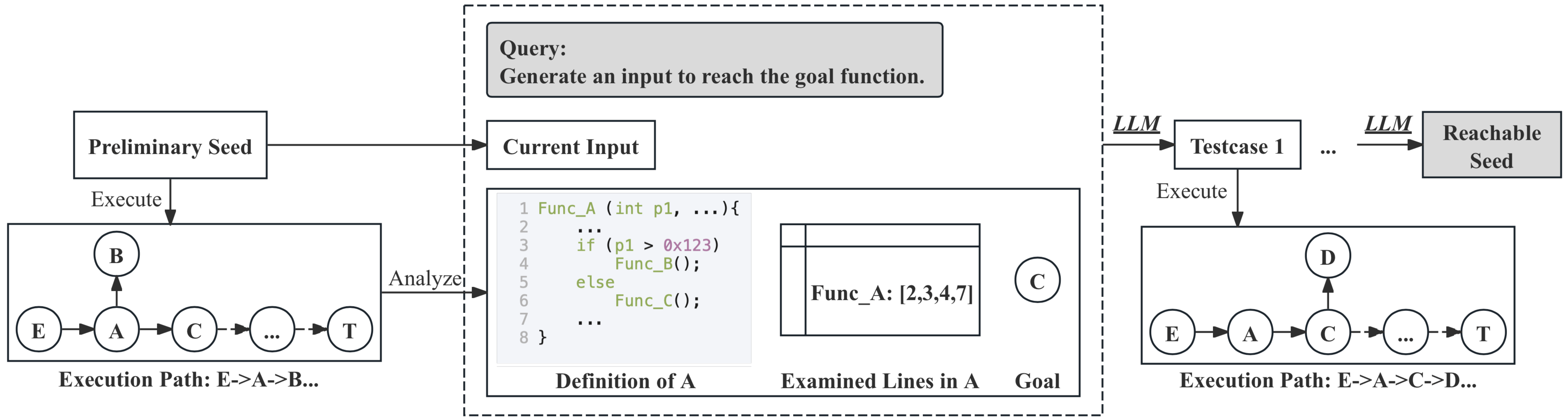}
    % \vspace{-7mm}
    \caption{Reasoning along FCC. The aim is to generate a seed that can reach the target vulnerable function \textit{T}.}
    \label{fig:reason_fcc}
\end{figure*}

\subsection{Reachable Seed Generation}

Seeds can impact the efficiency of directed fuzzing, motivating us to generate reachable seeds that can reach target locations.
For this purpose, it is essential to have an understand of the target program.
Reachable seeds must satisfy program logic to examine target code regions. For example, when testing a specific feature of image format conversion in an image processing program, an image file is more suitable as a seed than a text file.
Our \tool{} first creates preliminary seeds that at least meet the format requirements of the target program.
Then, if a complete Function Call Chain (FCC), which contains execution paths from the entry point to the target vulnerable function, can be obtained through static analysis, \tool{} employs LLMs to reason about reachable seeds along the FCC.
If not, \tool{} uses LLMs to infer reachable seeds based on the functionalities of neighbor functions.

\subsubsection{Preliminary Seed Generation}

Our goal is to continuously optimize the seeds until their execution paths reach the target location. Specifically, we aim to create preliminary seeds that exercise execution paths as close as possible to the location. 
Importantly, when \tool{} generates preliminary seeds, it also generates the associated command options. 
Command options significantly impact the effectiveness of directed fuzzing; if an option is unrelated to the target location, no input can successfully examine target buggy code regions.
Therefore, we begin by comparing the functionality of the target location with the \textit{Program Usage} information to identify which command options should be used to accept program input with the LLM.
In our query to the LLM, we set the \textit{Task} to analyze which program command is most likely to activate the target function, using the \textit{Answer Template} to obtain the appropriate command and descriptions. 
In addition, we provide the \textit{Function Summary} of the vulnerable function and the \textit{Program Usage} information as the \textit{Attachment} for LLM context.
Finally, %based on this command and its description, we ask 
the LLM generates the corresponding preliminary seed.
For the later fuzzing process, we will fixate the program command but mutate the seed.

For example, when generating preliminary seeds for the bug CVE-2017-16828 in the program \texttt{readelf}, we first obtain the \textit{Program Usage} through the LLM. Then, we analyze the vulnerable function \texttt{display\_debug\_frames()} and find that it processes and displays the contents of DWARF debugging frame sections, specifically the \texttt{.debug\_frame} and \texttt{.eh\_frame}.
After providing the \textit{Function Summary} and \textit{Program Usage} to the LLM for analysis, it identifies the command most likely to reach the vulnerable function as \texttt{readelf --debug-dump=frames file.elf}. The command description indicates that \texttt{file.elf} should be a valid ELF binary compiled with debugging symbols (\eg, using the \textit{-g} flag in GCC).
We then ask the LLM to generate a preliminary seed, which is an ELF file matching this description, to replace the placeholder \texttt{file.elf}.

\begin{figure*}
    \centering
    \includegraphics[width=\linewidth]{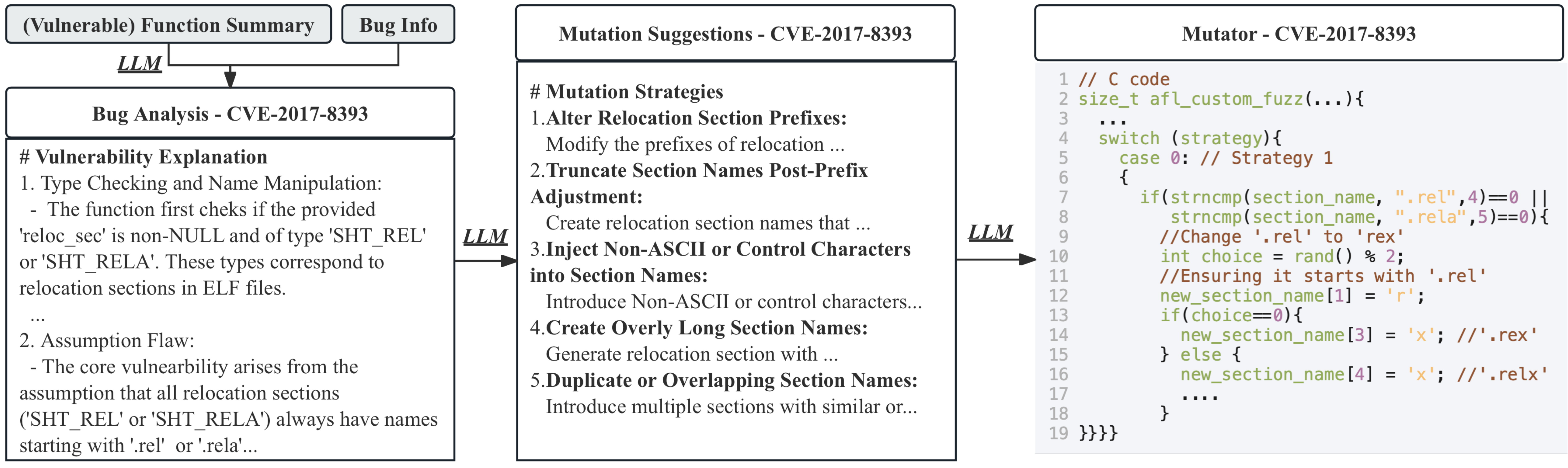}
    % \vspace{-7mm}
    \caption{Bug-specific Mutator. Based on LLM, \tool{} generates code to perform mutations.} 
    \label{fig:mutator}
\end{figure*}

\subsubsection{Reasoning Along Function Call Chain}\label{subsec:reason_fcc}

Relying solely on the summary of the target vulnerable function for seed generation is often insufficient to reach the target. This limitation arises because the execution path from the program's entry point to the target function may pass through multiple intermediate functions that are not considered in seed generation. Consequently, the requirements imposed by these intermediate functions on the seed are overlooked, which can cause the generated seed to deviate along incorrect paths.
To address this, \tool{} leverages LLMs to adjust the preliminary seed along an FCC, if a complete FCC that connects entry point and target function is available.

Overall, for this purpose, \tool{} takes two steps.
First, it uses Clang AST analyzer to get the complete FCC. Next, it gradually generates a reachable seed by analyzing deviation functions, which deviate the execution from the vulnerable FCC. 
For example, as shown in Figure~\ref{fig:reason_fcc}, to generate a seed that can reach the target vulnerable function \textit{T}, while the preliminary seed exercises the execution path $E\rightarrow A\rightarrow B$, which deviates from the target \textit{T}. 
Since functions $E$ and $A$ are both present in the complete FCC, \tool{} analyzes which one is the deviation function. 
By calculating the distances between functions $E$ and $T$ and between $A$ and $T$, \tool{} identifies function $A$ as the deviation function because it is closer to the target function $T$. 
The distance is defined as the number of edges in the shortest path from the source function to destination function. 
As a result, \tool{} obtains the definition of $A$, the lines examined by the preliminary seed, as well as the next function after the deviation function $A$ along the complete FCC, which, in this case, is function $C$. 
% shown in Figure~\ref{fig:reason_fcc}.
With this set of information, \tool{} can query LLMs to generate a testcase to reach the next function $C$ (the goal). 
%\tool{} uses the information of current input, the definition of deviation function $A$, the examined lines in the function, and the goal as the context.
Specifically, we set the \textit{Task} in the query as \textit{``Based on the provided function body and the current input that leads to a specific execution path, how should the input be modified to guide the program to reach the target function?''}. 
% The query includes the definition of function $A$, the current input, its corresponding execution path, and the target location as context.
After querying the LLM and receiving a response, we adjust the input accordingly and execute it, checking the execution path of the modified input. If the modified input reaches function $C$, \tool{} uses the corresponding testcase as the current seed, and repeat the above process to generate a new testcase that can reach the next goal.
This continues until \tool{} successfully generates a seed that can reach the target vulnerable function.
If not, we continue querying the LLM for further seed optimization guidance.

Notably, directly asking LLMs to generate an input meeting all code branches and reaching the optimization target is impractical.
This limitation arises because our prompt includes only a single function definition (the function $A$ in the example), whereas many branch conditions within this function depend on other functions or macro-defined parameters. 
%Without their code definitions, the LLM's analysis is limited. 
To address this, we inform LLMs of this limitation within our prompt, asking them to infer the potential roles of functions or parameters based on their names.
% \zhu{Do we find the definitions of other functions and use LLM to analyze them?}. 
As a result, multiple seeds can be generated regarding these potential roles, 
% each attempting different values in the input to navigate through branches and satisfy conditions.
aiming to satisfy path constraints. 
For example, when optimizing the seed for the program \texttt{cjpeg}, a path constraint compares a variable named \texttt{maxval} with a macro-defined value \texttt{MAXJSAMPLE}. Since only the relevant function definition is provided in the LLM query, without the macro definition, the exact value of \texttt{MAXJSAMPLE} is unknown. However, the LLM can infer that \texttt{MAXJSAMPLE} likely represents the maximum color component value in an image. Based on this, it assumes \texttt{MAXJSAMPLE} could be 255 (for 8-bit images) or 4095 (for 12-bit images) and generates two different seeds to test these possibilities.
This seed generation process continues iteratively until a generated seed either reaches the optimization target or progresses to a function closer to the target.

\subsubsection{Reasoning Based on Functionality} \label{sec:broken_fcc}

Static analysis may fail to produce a complete FCC due to program complexities such as nested macro definitions and function calls using function pointers and memory addresses. 
Many directed fuzzers face similar challenges, leading to incomplete calculations of distances between current execution blocks and target code blocks~\cite{bohme2017directed, chen2018hawkeye}.
To address this issue, we leverage the summarization capabilities of LLMs to generate reachable seeds. 
By analyzing the neighboring functions - those that can reach the target function - \tool{} gains insights into the target function's role within a broader program context. 
%Neighboring functions are those that can reach the target function.
Our insight is that, if a complete FCC cannot be obtained from the entry point, an incomplete FCC may still be derived by starting from the neighboring functions of the vulnerable one. LLMs can then be used to generate inputs that reach these neighboring functions based on their functionalities. 
% Once we identify inputs that reach neighboring functions, we can apply the method described in Section~\ref{subsec:reason_fcc} to generate a reachable seed.

\tool{} first uses static analysis tools (\eg, Clang) to identify all neighboring functions of the target function. It then randomly selects a neighboring function and queries LLMs about the program options and inputs that can lead to this function, with the context of \textit{Program Usage} and \textit{Function Summary}. 
By analyzing the instrumentation execution logs, we can determine whether these inputs reach the neighboring function. If they do not, we randomly select another neighboring function and repeat the input generation process based on the functionality of the newly selected function.
Finally, if an input successfully reaches a neighboring function, we apply the method outlined in Section~\ref{subsec:reason_fcc} to generate reachable seeds.

\input{table/benchmarks}

% \vspace{-3mm}
\subsubsection{Issues for Input Generation Process} \label{sec:input_generation}
The first issue is the input format that impacts how we query LLMs. 
Input formats can range from simple strings to more complex data types, such as images or ELF files. For string-type inputs, we query LLMs to directly generate the string. For complex inputs, we query LLMs to generate code that can produce these inputs, as they  cannot be directly generated by LLMs. For example, we may query as \textit{Please provide a Python script that, when executed, outputs a file that meets the requirements.}
Another issue is that the generated code may not be runnable. To ensure the reliability of codes generated by LLMs, we feed any error messages from code execution back into LLMs, allowing them to correct the code. 
This process is repeated until the code executes without errors.
The final issue is that, during the process of guiding seeds toward the target vulnerable function, we may fail to generate a seed that reaches the target, stalling the testing process. 
To address this, our prototype imposes a time limit for this phase. If no seed reaches the target within a specific time duration, such as one hour, we select a subset of the generated seeds to serve as seeds for fuzzing.
% for the subsequent steps.

% \vspace{-4mm}
\subsection{Bug-Specific Mutators}

As shown in Figure~\ref{fig:workflow}, merely reaching the target locations does not necessarily trigger bugs, as specific execution states must also be met. To reduce randomness in mutating reachable seeds, \tool{} employs LLMs to generate bug-specific mutators, tailoring mutation schemes to target particular bugs. 
As shown in Figure~\ref{fig:mutator}, we begin by using the \textit{Bug Info} extracted from the bug report, along with the \textit{Function Summary} of the vulnerable function, to query the LLM. The \textit{Task} of this query prompts the LLM to provide a \textit{Bug Analysis} report that summarizes the bug's cause and how to trigger the vulnerability.
Next, we ask the LLM to generate a fuzzing mutation strategy based on the \textit{Bug Analysis} report. In this query, we include examples of mutation strategies and their explanations in the \textit{Attachment} and \textit{Suggestion} sections, respectively, to guide the LLM in generating effective new mutation strategies.
Finally, these new mutation strategies are sent to the LLM with a \textit{Task} to translate them into C language code. To ensure code quality, real C language mutator code examples and explanations are included in the query context. If the generated code fails to compile, we ask the LLM to revise the code based on the original mutator code and compilation error messages, and attempt compilation with the updated code.

Additionally, some hidden or deeper code errors may not appear during compilation but could cause the fuzzer using the mutator to crash or experience significantly reduced execution efficiency. To ensure the generated mutator code runs correctly, we conduct a trial run before executing the full fuzzing process, checking for any issues in the mutator code.
In this trial, we verify that the fuzzing program runs smoothly and that specific fuzzing metrics (such as executions per second) fall within expected ranges. If crashes or abnormal metrics occur, we request the LLM to regenerate a replacement mutator.
Meanwhile, in the directed fuzzing phase, the mutation strategies generated by the LLM are often deterministic. Repeatedly using the same set of strategies can result in wasted time on redundant test cases. To ensure a diverse and effective testing, we request the LLM to generate a new set of mutation strategies every hour.

%% file: table/benchmarks.tex
% Please add the following required packages to your document preamble:
% \usepackage{multirow}
\begin{table*}[!t]

\centering

\caption{Benchmark.}

% \begin{adjustbox}{width=\textwidth,center}
\resizebox{0.95\linewidth}{!}{
\begin{threeparttable}

\begin{tabular}{cl|llcl}
\toprule
\hline
\textbf{Program} &
  \textbf{CVE ID} &
  \multicolumn{1}{c}{\textbf{Vulnerability Type}} &
  \multicolumn{1}{c}{\textbf{Vulnerable Location}} &
  \textbf{Indirect Calls}\tnote{1} &
  \multicolumn{1}{c}{\textbf{Edit}\tnote{2}} \\ \hline
\multirow{2}{*}{cjpeg}   & CVE-2018-14498 & Heap-Buffer Overflow     & rdbmp.c:get\_8bit\_row                            & T &         N             \\
                         & CVE-2020-13790 & Heap Buffer Overflow     & rdppm.c:get\_rgb\_row                             & T &  N                    \\ \hline
\multirow{3}{*}{cxxfilt} & CVE-2016-4487  & NULL Pointer Dereference & cplus-dem.c:register\_Btype                       & F & Y    \\
                         & CVE-2016-4489  & Integer Overflow         & cplus-dem.c:string\_appendn                       & F &   N                   \\
                         & CVE-2016-4491  & Stack Overflow           & cp-demangle.c:d\_print\_comp\_inner               & T &  N                    \\ \hline
\multirow{3}{*}{objcopy} & CVE-2017-8393  & Global Buffer Overflow   & elf.c:\_bfd\_elf\_get\_reloc\_section             & T &         N             \\
                         & CVE-2017-8394  & NULL Pointer Dereference & objcopy.c:filter\_symbols                         & F & Y    \\
                         & CVE-2017-8395  & NULL Pointer Dereference & compress.c:\_bfd\_generic\_get\_section\_contents & T & Y \\ \hline
\multirow{3}{*}{objdump} & CVE-2017-8392  & Heap Buffer Overflow     & dwarf2.c:parse\_comp\_unit                        & T &          N            \\
                         & CVE-2017-8397  & Heap Buffer Overflow     & reloc.c:bfd\_perform\_relocation                  & T & Y    \\
                         & CVE-2018-17360 & Heap Buffer Overflow     & peigen.c:pe\_print\_edata                         & T & N                     \\ \hline
strip                    & CVE-2017-7303  & NULL Pointer Dereference & elf.c:section\_match                              & T &       N               \\ \hline
nm                       & CVE-2017-14940 & NULL Pointer Dereference & dwarf2.c:scan\_unit\_for\_symbols                 & T &      N                \\ \hline
readelf                  & CVE-2017-16828 & Heap Buffer Overflow     & readelf.c:display\_debug\_frames                  & T &       N               \\ \hline
% xmlint                  & CVE-2024-34459 & Heap Buffer Overflow     & xmllint.c:xmlHTMLPrintFileContext                 & F &        N              \\ \hline
\end{tabular}

% \end{adjustbox}

\label{tab:benchmark}

\begin{tablenotes}
\footnotesize

\item[1] \textbf{Indirect Calls}: Indirect calls lead to incomplete function call chain from the entry function to the target vulnerable function. If there is an indirect call on the function call chain, it will be marked as \textit{T}; otherwise it is \textit{F}.

\item[2] \textbf{Edit}: Some CVE reports do not directly state the vulnerable function names or provide incorrect ones. Therefore, we manually edit reports to add or modify the vulnerable function names in such CVE reports, and mark it as \textit{Y}; otherwise, we mark it \textit{N}.
% \item[2] \textbf{Comments}: The CVE reports of some projects do not directly state the vulnerable function names required for \tool{} analysis or provide incorrect names. Therefore, we manually add or modify the vulnerable function names in such CVE reports.
\end{tablenotes}
\end{threeparttable}
}

\end{table*}

%% file: body/eval.tex
\section{Evaluation} \label{sec:eval}

\newcommand{\rqone}{\textbf{RQ1: Preparation Time for Fuzzing}}
\newcommand{\rqtwo}{\textbf{RQ2: Efficiency of Reachable Seeds}}
\newcommand{\rqthree}{\textbf{RQ3: Efficeincy of Bug Detection}}
\newcommand{\rqfour}{\textbf{RQ4: LLM Model Influence}}
\newcommand{\rqfive}{\textbf{RQ5: Reasoning Ability of LLM}}

In this section, we address the following research questions through various experiments:

\begin{itemize}[left=0pt]

    \item \rqone~- Does \tool{} require more time for fuzzing preparation?

    \item  \rqtwo~- Do the reachable seeds generated by \tool{} result in more efficient bug detection than other seeds?
    
    \item  \rqthree~- Does \tool{} perform better than other directed fuzzers in detecting various vulnerabilities?
    
    \item  \rqfour~- Do different GPT models affect \tool{}? 

    % \item  \rqfive~- Does the LLM already have the knowledge of CVE reports or output bug information via its reasoning capabilities?
    % Is the LLM leveraging the CVE knowledge in the knowledge base to assist \tool{} rather than relying on its reasoning capabilities?

\end{itemize}

%========================= Enviroment =========================
\noindent\textbf{Experiment Environment.}
% We conducted our experiments on a machine equipped with two Intel(R) Xeon(R) Gold 6138 CPUs at 2.00GHz, which includes a total of 40 CPU cores and 80 CPU threads. Using Ubuntu 20.04 LTS, each fuzzing session ran on a Docker container with a single CPU thread and 2GB of RAM allocated. We used all 80 CPU threads, all running the same fuzzing session.
Our experiments were conducted on a machine equipped with two Intel(R) Xeon(R) Gold 6138 CPUs at 2.00 GHz, providing a total of 40 CPU cores and 80 threads. Running Ubuntu 20.04 LTS, each fuzzing session was executed within a Docker container with one CPU thread and 2GB of RAM allocated per session. We utilized all 80 CPU threads, each running an instance of the same fuzzing session.

\input{table/Preparation}

%========================= Prototype Implementation =========================
\noindent\textbf{Tool Implementation.}
% We implemented \tool{} with  approximately \todo{xxx} lines of Python code.
% We developed our \tool{} based on AFL++~\cite{AFLplusplus20} because it supports user-defined mutators. We integrate our LLM-generated mutators into AFL++ so that our \tool{} has both bug-specific mutators and random mutators.
% Our \tool{} can automatically perform the fuzzing process with the information of bug report, project files, and compilation commands.
% We use GPT-4o as the LLM and use OpenAI's APIs (2024-08-06 API) to interact with it. 
% The static analysis tool we use in the methodology is the Abstract Syntax Tree (AST) analysis tool from Clang\footnote{https://clang.llvm.org/docs/IntroductionToTheClangAST.html}. 
We implemented \tool{} with approximately 2400 lines of Python code. Our tool is built upon AFL++~\cite{AFLplusplus20}, which supports user-defined mutators, enabling us to integrate LLM-generated mutators alongside AFL++'s random mutators. This allows \tool{} to leverage both bug-specific and random mutation strategies.
% \tool{} automates the fuzzing process using information such as bug reports, project files, and compilation commands. 
We use GPT-4o as the LLM and interact with it via OpenAI's APIs (2024-08-06 API). The static analysis component in our methodology employs Clang’s Abstract Syntax Tree (AST) analysis tool\footnote{https://clang.llvm.org/docs/IntroductionToTheClangAST.html}.

% \todo{implementation details \begin{itemize}
%     \item We implemented the prototype using approximately \todo{xxx} lines of Python code. After specifying the contents of the CVE report, the project files, and the compilation commands, the prototype is able to automatically perform various analyses and initiate directed fuzzing.
%     \item We chose GPT-4o (2024-08-06 API) as the large language model for our methodology. In the prototype, we integrated the use of OpenAI's model API. When the need arises to call upon the large language model for analysis, the prototype will invoke pre-stored prompts, fill in the corresponding context, and send the query via the API.
%     \item To better interact with the prototype and maintain a lightweight approach, we adopted Clang's Abstract Syntax Tree (AST) analysis tools for our static code analysis.
%     \item We have chosen AFL++ as our fuzzer. AFL++ offers a wide range of parameters to finely tune its functionality and also supports the use of custom mutation algorithms by testers. AFL++ supports using its native Havoc algorithm in conjunction with user-defined mutators. Therefore, in the custom mutator, we can request the large language model to focus more on trying specific values and combinations, leaving the fully random mutations to be handled by AFL++.
% \end{itemize}}

%========================= Baseline =========================
\noindent\textbf{Baseline Fuzzers.} We select four state-of-the-art directed fuzzers as baselines to evaluate \tool{}. 
% In the following experiments, we selected the following directed fuzzers:

\begin{itemize}[left=0pt]

    \item AFLGo~\cite{bohme2017directed} is the first and commonly-used fuzzer to be compared. It uses distance to guide fuzzing towards target locations. It uses mutators from AFL, whose mutators are pre-determined and not bug-specific. 

    \item Beacon~\cite{huang2022beacon} only exercises execution paths that can reach target locations so that it focuses on exploring relevant code. It is developed upon AFLGo, indicating that it uses mutators from AFL.

    \item WindRanger~\cite{du2022windranger} guides executions towards target locations based on deviation basic blocks, which deviate the execution paths from target locations. It is developed based on AFL, indicating that it uses mutators from AFL.

    \item SelectFuzz~\cite{luo2023selectfuzz} instruments code regions that are relevant to target locations so that it can focus on exploring the relevant code regions. It is also developed based on AFL, indicating it uses mutators from AFL.
    
\end{itemize}

%========================= Benchmarks =========================
\noindent\textbf{Benchmarks.} 
Table~\ref{tab:benchmark} lists all the vulnerabilities involved in our experiments, as well as the details of them. 
We select 14 vulnerabilities from 8 different programs, including \texttt{cjpeg}, \texttt{cxxfilt}, \texttt{objcopy}, \texttt{objdump}, \texttt{strip}, \texttt{nm}, and \texttt{readelf}.
The 14 vulnerabilities have diverse types of bugs including heap-buffer overflow, NULL pointer dereference, integer overflow, stack overflow, and global buffer overflow.
These vulnerabilities are used to evaluate fuzzing in the papers of baseline fuzzers.
To ensure a fair comparison, we adopt the same compilation options and commands in the evaluation. 
In our evaluation, all experimental tests are repeated 4 times. 
We present the median value in the tables.
% To avoid the impact of extreme values and to facilitate some validation calculations, the final value for each experiment will be taken as the median of the four runs.

% To ensure the fairness of the test results, we select a subset of the test suite used in previous work~\cite{tae2023DAFL} (some programs had conflicts with the compilation commands and testing tools, i.e. Address Sanitizer) as our test set (14 vulnerabilities). Furthermore, to minimize irrelevant interference, we adopted the same compilation options and commands used in \cite{tae2023DAFL} work to set up our test environment. 
% In addition, to demonstrate the reasoning capabilities of the LLM, we also selected a recently disclosed CVE project as a test subject.

%========================= Value =========================
% \noindent\textbf{Data Value.} In our evaluation, all experimental tests will be repeated \textbf{four} times. To avoid the impact of extreme values and to facilitate some validation calculations, the final value for each experiment will be taken as the median of the four runs.

%================================================================
% \subsection{RQ1: Preparation Time for Fuzzing}
\subsection{\rqone}

\tool{} has four steps that prepare for later fuzzing process, including static analysis for analyzing a program's function call chain (\textit{SA}), program option construction using RAG (\textit{RAG}), seed optimization to generate reachable seeds (\textit{Opt}), and generation of bug-specific mutators (\textit{Mutator}).
% We divide the preparation work for \tool{} into four parts: Static analysis for analyzing the program's function call relationships (SA), Constructing program inputs using RAG (RAG), Optimizing program inputs to reach the target location (Opt), and Generating the fuzzing Mutation strategy (Mutator). 
Thus, in this evaluation, we compare the preparation time across fuzzers to check whether \tool{} significantly increases the preparation time for directed fuzzing.
As shown in Table~\ref{tab:preparation}, 
% on average, the process \textit{Opt} spends the most time, with 53\% more time than the second process \textit{RAG}.
on average, the \textit{Opt} process takes the most time, consuming 53\% more time than the second longest process, \textit{RAG}.
This is because \textit{Opt} process involves repeated interactions between querying the LLMs to generate inputs and executing the target program to obtain feedback. 
% According to Table~\ref{tab:preparation}, the processing time of \textit{Opt} required by different vulnerabilities varies, from 72 seconds to 957 seconds. This discrepancy is because the target locations have different depths, which indicate the complexity of control flow and data flow to reach the target.
% Too deep targets may lead to timeout, meaning \tool{} cannot generate the reachable seed within a limited time.
%As shown in Table~\ref{tab:preparation}, 
The processing time for \textit{Opt} varies significantly across different vulnerabilities, ranging from 72 seconds to 957 seconds. This variation occurs because target locations differ in depth, reflecting the complexity of the control flow and data flow needed to reach the target. Excessively deep targets may result in timeouts, preventing \tool{} from generating a reachable seed within the allotted time.
% \textit{RAG} takes the second longest time because it involves many operations on slicing and storing text and code, which positively correlates with the size
% of the target program.
\textit{RAG} takes the second longest time due to the operations required for slicing, encoding, and retrieving text and code, which are positively correlated with the size of the target program. 
% The process of generating fuzzing mutation strategies in \tool{} is relatively fixed, ranging from 83 seconds to 147 seconds. 
The process \textit{Mutator} for generating fuzzing mutation strategies in \tool{} is relatively consistent, taking between 83 and 147 seconds.
% This is mainly because this process takes input as reports of vulnerabilities and function summaries that are already obtained from previous stages.
This is primarily because this process uses vulnerability reports and function summaries that are gathered in previous stages as input.
Finally, the processing time for \textit{SA} depends on the size of target programs, and it takes the least time, with 47 seconds on average.

\input{table/Ablation_Seed}

% Overall, compared to other fuzzers, \tool{} spends less time for the total four steps than AFLGo, Beacon, and SelectFuzz, but more time than WindRanger.  
Overall, compared to other fuzzers, \tool{} requires less time across the four steps than AFLGo, Beacon, and SelectFuzz, but more time than WindRanger.
On average, \tool{} spends 62.6\%, 86.2\%, and 60.1\% less time than AFLGo, Beacon, and SelectFuzz (with TS), respectively. 
AFLGo spends considerable time on calculating  distances between basic blocks.
In addition to distance calculation, Beacon requires extra time for reachability analysis.
SelectFuzz also has a lengthy preparation when the Temporal-Specialization (TS) mode is enabled, as this mode analyzes indirect calls to obtain a potentially complete control flow graph.
This shows the advancement of our \tool{}, which leverages LLMs to analyze functionalities instead of analyzing indirect calls.
WindRange spends the least time on preparation because its distance calculation is performed during the fuzzing process.
% Therefore, \tool{} is competitive in preparing fuzzing.
Therefore, \tool{} demonstrates competitive performance in fuzzing preparation.

%CVE-2018-17360 Timeout problem.

%================================================================
% \subsection{RQ2: Seed Comparison}
\subsection{\rqtwo}

In this section, we evaluate the efficiency of \tool-generated reachable seeds, comparing with other sets of seeds. 
Since seed generation and mutators are tightly coupled in \tool{}, we do not evaluate \tool{} on different sets of seeds. 
The four directed fuzzers are evaluated on three sets of initial seeds, where each fuzzer runs 24 hours on each vulnerability.
As shown in Table~\ref{tab:ablation_seed}, the three sets of seeds include Original Seeds (OS), Simple Seeds (SS), and \tool-generated Seeds (RLS).
Original seeds are the set of seeds collected by Kim \etal~\cite{kim2024evaluating}.
Simple seeds are the set of naive seeds that only meet the minimum input requirements and do not aim to trigger any specific functionality. For example, in the case of the program \texttt{cxxfilt}, which takes a character-type function symbol name as input, we might provide a simple test seed like \textit{\_Z3foov}.
\tool-generated seeds are the set of seeds generated by \tool{}.

Table~\ref{tab:ablation_seed} shows the time to bugs for different sets of initial seeds. 
% \tool-generated seeds take the least time to expose bugs when running all the four fuzzers. 
% Equipped with \tool-generated seeds, all the four fuzzers take the least time to expose bugs, with AFLGo on 11 bugs, Beacon on 5 bugs, WindRanger on 9 bugs, and SelectFuzz on 4 bugs.
With \tool-generated seeds, all four fuzzers achieve the fastest bug exposure time. 
% Specifically, AFLGo with RLS exposes bugs the most quickly on 10 bugs, Beacon on 6, WindRanger on 9, and SelectFuzz on 6, bringing the total number to 31.
Specifically, AFLGo with RLS is the fastest at exposing 10 bugs, Beacon at 6, WindRanger at 9, and SelectFuzz at 6, bringing the total number to 31.
The second fastest seed set is the Original Seeds, with AFLGo performing the best on 3 bugs, Beacon on 3, WindRanger on 2, and SelectFuzz on 3.
% On average, when ruling out the bugs with timeouts (CVE-2018-17360 and CVE-2017-14940), AFLGo with RLS discovers bugs 4.8x faster than OS, and 4.3x faster than SS.
On average, excluding the bugs with timeouts (CVE-2018-17360 and CVE-2017-14940), AFLGo with RLS discovers bugs 4.8$\times$ faster than OS and 4.3$\times$ faster than SS.
Similarly, Beacon with RLS exposes bugs 3.4$\times$ faster than OS and 2.25$\times$ faster than SS.
WindRanger with RLS exposes bugs 2.1$\times$ faster than OS and 4.3$\times$ faster than SS.
% As for SelectFuzz, SelectFuzz with RLS exposes bugs 0.4$\times$ slower than OS but 1.8$\times$ faster than SS.
On average, SelectFuzz with RLS discovers bugs slower than OS and SS; but on 6 bugs, SelectFuzz with RLS is the fastest to expose bugs. 
The drop of average speed mainly because it takes 17273 seconds to expose CVE-2017-7303.
On some bugs, \tool-generated seeds can help fuzzers discover bugs within tens of seconds, more than 1,000$\times$ faster than other seed sets.
For example, on the bug CVE-2018-14498, AFLGo with RLS discovers the bug 1,740$\times$ faster than OS.
This demonstrates that our \tool-generated seeds can significantly improve the efficiency of bug exposure.

\subsection{\rqthree}

\input{table/Ablation_Mutation}

\input{table/Ablation_LLM_Model}

In this section, we evaluate the performance of different fuzzers. To ensure a fair comparison, all fuzzers are equipped with \tool-generated reachable seeds, running for 24 hours.
Table~\ref{tab:ablation_mutation} shows that, overall, \tool{} performs the best, discovering 7 bugs the most quickly.
For the rest 6 exposed bugs, although \tool{} does not expose bugs the most quickly, the time to exposing bugs is close to the fastest one.
For example, Beacon discovers CVE-2016-4487 in 3 seconds while \tool{} exposes it in 5 seconds.
Of the 13 bugs that \tool{} can expose, 8 are discovered within 60 seconds, and 2 within a few hundred seconds, showing the efficiency of our \tool{}.
% The speedup of bug exposure by \tool{} ranges from  
% Although we provide the \tool-generated reachable seeds, the speedup of bug exposure by \tool{} can still be up to 3018.8$\times$ than Beacon.
% Even with \tool-generated reachable seeds provided, \tool{} can still achieve a bug exposure speedup ranging from 1.01$\times$ to 2.7$\times$, compared to the second fastest exposure.
Even with \tool-generated reachable seeds provided, \tool{} achieves a bug exposure speedup ranging from 1.01$\times$ to 2.7$\times$ compared to the second-fastest exposure.
% The most speedup is when exposing CVE-2016-4491, where \tool{} only takes 2 seconds to expose the bug while SelectFuzz takes 13114 seconds to expose it, 67057$\times$ slower than \tool{}.
The greatest speedup is observed with CVE-2016-4491, where \tool{} takes only 2 seconds to expose the bug, while SelectFuzz requires 13,114 seconds, making it 6557$\times$ slower than \tool{}.
Since all fuzzers use \tool-generated seeds, the superiority of \tool{} must stem from its advanced mutation strategy.
Therefore, both the reachable seed generation and bug-specific mutator construction are effective and efficient in exposing bugs.

%================================================================
% \subsection{RQ4: Model Choices} 
\subsection{\rqfour}
\label{eval:RQ4}

In this section, we evaluate the performance of different LLM models, including GPT-4 (GPT-4-0125-preview, GPT-4o, GPT-o1-mini (GPT-o1-mini-2024-09-12), and GPT-o1-preview (GPT-o1-preview-2024-09-12).
We assess LLM models on three tasks, including command option construction via RAG (\textit{RAG}), optimization for reachable seed generation (\textit{Opt}), and bug-specific mutator generation (\textit{Mutator}). 
Table~\ref{tab:llm_model} presents the time each GPT model makes to assist \tool{} in these tasks.
Updates in GPT models do not generally reduce completion time.
GPT-o1-preview-2024-09-12 is newer than GPT-4o, but it completes the \textit{Opt} task slower than GPT-4o due to higher computational costs.

As shown in Table~\ref{tab:llm_model}, for the task \textit{RAG}, GPT-4o and GPT-o1-mini perform similarly in this task, with GPT-4 slightly slower, while GPT-o1-preview takes significantly longer due to slower query response time.
For the task \textit{Opt}, GPT-4o proves more efficient than GPT-o1-mini, while GPT-4 and GPT-o1-preview are considerably slower, with timeouts in some cases. GPT-o1-mini shows faster single-query responses than GPT-4o but sometimes requires re-queries due to formatting inconsistencies. GPT-4 occasionally produces incomplete or incorrect code analyses for complex programs, leading to inefficiencies. GPT-o1-preview, though slower per query, provides high-quality responses, reducing the total number of queries needed.
For the task \textit{Mutator}, GTP-4, GPT-4o, and GPT-o1-mini perform similarly in completion time, with GPT-o1-preview slightly slower. However, GPT-o1-mini and GPT-o1-preview frequently make errors in code generation, such as undefined variables in C code or indentation issues in Python. This requires additional LLM queries to correct the code.

After the comprehensive evaluation, GPT-4o emerges as the best choice for \tool{}. It consistently performs well in seed optimization and code analysis, avoiding formatting issues and providing efficient task completion, unlike GPT-o1-mini, GPT-4, and GPT-o1-preview. GPT-4o effectively balances accuracy and efficiency, enhancing \tool{}'s overall performance.

\subsection{Case Study for the Limitations of \tool}\label{subsec:cases}

% In this section, we analyze two cases about the challenges faced by \tool{}.
In this section, we analyze two cases highlighting the challenges faced by \tool{}.
Table~\ref{tab:ablation_seed} shows that RLS performs worse than OS on a few CVE vulnerabilities, such as CVE-2018-17360 and CVE-2017-16828. 
We analyze the reasons in this section.
% For CVE-2018-17360, \tool{} cannot generate reachable seeds due to the complexity of constructing seeds, which further leads to timeouts in fuzzing.
% The OS can expose this bug may because such seeds are suitable for triggering the bug.
% For CVE-2017-14940, \tool{} can generate the reachable seeds but the buggy states are hard to satisfy, resulting that all fuzzers fail to trigger this bug.
% We will discuss the reason for the poor performance in detail in Section~\ref{subsec:cases}.

\begin{figure}
    \centering
    \includegraphics[width=0.75\linewidth]{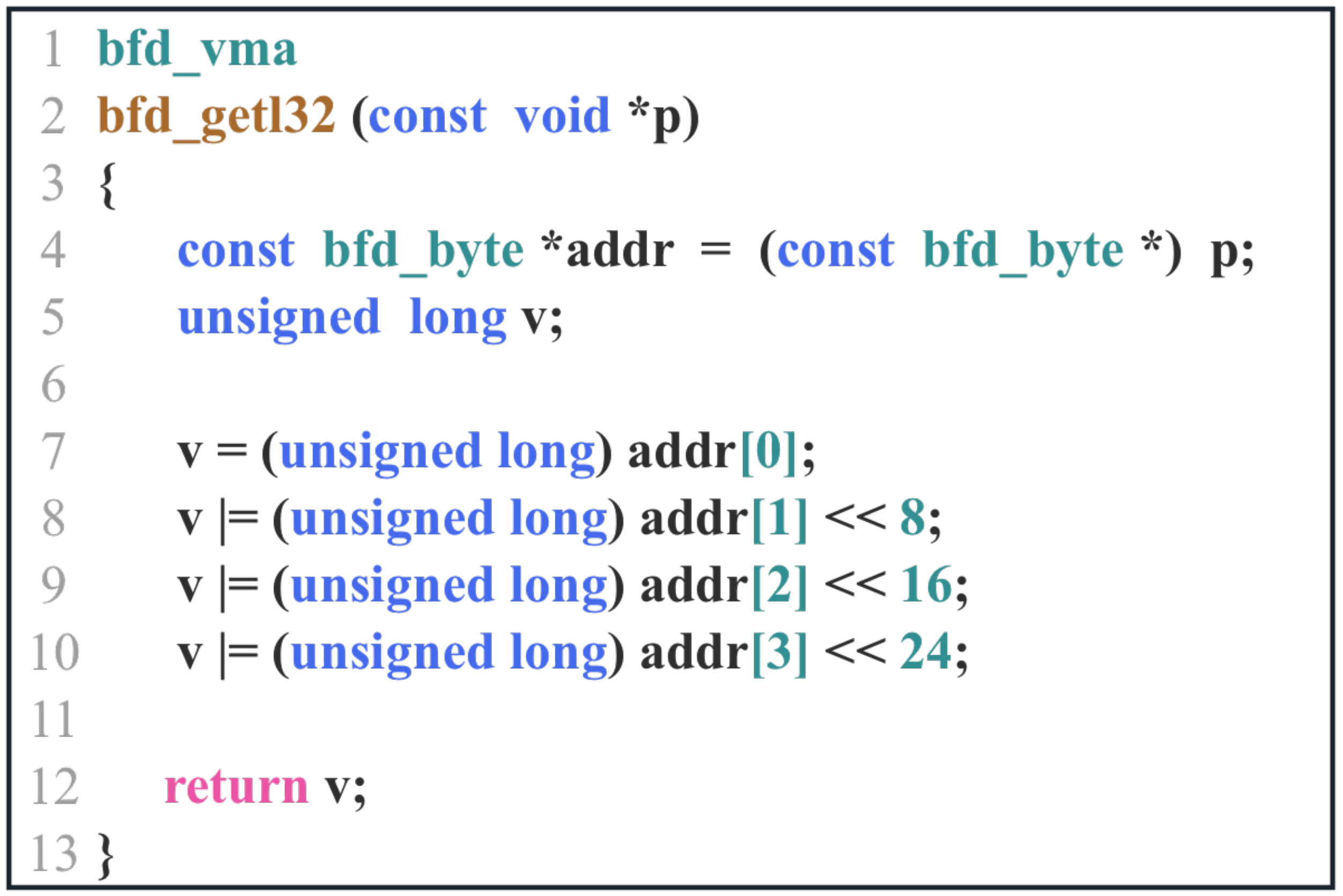}
    % \vspace{-5mm}
    \caption{Function body of \texttt{bfd\_getl32()}.}
    \label{fig:bfd_getl32}
\end{figure}

\subsubsection{Case 1: Lack of Neighbor Functions} \label{cve-2018-17360}

To address the issue of incomplete function call chains, \tool{} proposes to generate seeds based on the functionalities of neighbor functions, which improves the possibility of generating reachable seeds.
However, if we cannot obtain any neighbor function of the target function, we have to generate reachable seeds based on the target function.
This may lead to the failure of reachable seed generation.
For example, Figure~\ref{fig:bfd_getl32} shows the target vulnerable function as described in the bug report of CVE-2018-17360.
Clang fails to get both the complete function call chain and any neighbor functions of \texttt{bfd\_getl32()}. 
Therefore, in this case, \tool{} has to generate reachable seeds based on the only function \texttt{bfd\_getl32()}.
The functionality of this function is that it reads 4 bytes of data from a given memory address (\textit{p}) and combines them into a 32-bit unsigned integer.
Although such functionality is common in dealing with binary data, network protocols,
file format parsing, and hardware interactions, \tool{} still cannot successfully generate reachable seeds.
This is because LLMs have difficulties in identifying the functional module that \texttt{bfd\_getl32()} belongs to.
Thus, in our experiments, \tool{} is unable to complete the seed optimization for CVE-2018-17360 within one hour (Table~\ref{tab:preparation}). Due to its inability to generate a reachable seed, \tool{} is unable to trigger the vulnerability within 24 hours (Table~\ref{tab:ablation_mutation}).

In addition, in \texttt{Libjpeg} (\texttt{cjpeg}) related projects (CVE-2018-14498 and CVE-2020-13790), Beacon failed to find the complete function call chain from the entry function to the target function (\texttt{get\_8bit\_row()} for CVE-2018-14498 and \texttt{get\_rgb\_row()} for CVE-2020-13790), which led to the failure in identifying the target vulnerabilities.
Specifically, we analyzed the output results of Beacon's static analysis for both vulnerability projects.
In their distance calculation records, we found that Beacon incorrectly pruned parts of the target path because it could not resolve the call paths based on function pointers.
This indicates that Beacon's fuzzing cannot reach the target location.
As the result shown in Table~\ref{tab:ablation_seed} and Table~\ref{tab:ablation_mutation}, Beacon's static analysis fails to find the target and terminates after running for some time, and regardless of the seeds used, Beacon cannot trigger the bugs of these two CVEs in \texttt{Libjpeg}.

\subsubsection{Case 2: Limitation of LLM-Generated Seeds}

The reachable seeds generated by \tool{} significantly improve the efficiency of directed fuzzing.
However, such seeds introduce a new issue because LLMs may add too many details in the seeds.
Thus, in Table~\ref{tab:ablation_seed}, fuzzers equipped with \tool-generated seeds may trigger bugs slower than fuzzers with original seeds.
For example, LLMs add too many debug sections into the seeds for triggering the bug CVE-2017-16828 in the program \texttt{readelf}.
The CVE-2017-16828 describes a heap-based buffer over-read or crash issue that occurs when the program \texttt{readelf} processes a specially crafted ELF file containing debug information. Malicious DWARF data or incorrect DWARF offsets within the ELF file can lead to invalid pointer references or crashes. Since the vulnerability lies in parsing the input's (ELF file) debug information, \tool{} introduces various debug sections into the input during seed optimization to reach the target location. Although the generated seeds are capable of reaching the target, other sections of debug information are added into the seed, such as \texttt{.debug\_info}, \texttt{.debug\_line}, and \texttt{.eh\_frame}. 
The process also includes a substantial number of entries such as \texttt{.rela.debug\_info}, \texttt{.rela.debug\_aranges}, and \texttt{.rela.debug\_line}.
While these additional entries can potentially trigger the vulnerability, their volume inevitably increases the input length, which in turn reduces the efficiency of random mutations in finding the vulnerability.

%% file: table/Preparation.tex
\begin{table*}[!t]
\caption{Preparation Time of Each Fuzzer.}
\centering

% \begin{adjustbox}{width=\textwidth,center}
\resizebox{0.9\linewidth}{!}{
\begin{threeparttable}

\begin{tabular}{l|c|c|c|cc|ccccc}
\toprule
\hline
\multirow{2}{*}{\textbf{CVE ID}} &
  \multirow{2}{*}{\textbf{AFLGo}} &
  \multirow{2}{*}{\textbf{Beacon}} &
  \multirow{2}{*}{\textbf{WindRanger}} &
  \multicolumn{2}{c|}{\textbf{SelectFuzz}\tnote{1}} &
  \multicolumn{5}{c}{\textbf{\tool{}}\tnote{2}} \\ \cline{5-11} 
 &
   &
   &
   &
  \textbf{no TS} &
  \textbf{with TS} &
  \textbf{SA} &
  \textbf{RAG} &
  \textbf{Opt} &
  \textbf{Mutator} &
  \textbf{Total} \\ \hline
\rowcolorA 
CVE-2018-14498 & 256s  & 19s   & 37s & 47s   & 45s   & 24s & 95s  & 693s & 145s & 957s  \\
CVE-2020-13790 & 260s  & 19s     & 37s & 47s   & 49s   & 23s & 93s  & 423s & 121s & 660s  \\
\rowcolorA 
CVE-2016-4487  & 304s  & 3095s  & 77s & 152s  & 2063s & 48s & 193s & 177s & 126s & 544s  \\
CVE-2016-4489  & 307s  & 5068s  & 77s & 149s  & 1962s & 51s & 152s & 181s & 83s  & 467s  \\
\rowcolorA 
CVE-2016-4491  & 309s  & 6072s  & 76s & 149s  & 1870s & 45s & 178s & 100s & 147s & 470s  \\
CVE-2017-8393  & 1858s & 4852s  & 68s & 187s  & 2866s & 48s & 145s & 175s & 112s & 480s  \\
\rowcolorA
CVE-2017-8394  & 1846s & 5971s  & 70s & 187s  & 2543s & 50s & 162s & 157s & 93s  & 462s  \\
CVE-2017-8395  & 1853s & 4966s  & 68s & 183s  & 2555s & 54s & 191s & 150s & 87s  & 482s  \\
\rowcolorA
CVE-2017-8392  & 1991s & 8046s  & 88s & 171s  & 2464s & 49s & 279s & 417s & 89s  & 834s  \\
CVE-2017-8397  & 2016s & 9910s  & 89s & 166s  & 2355s & 51s & 287s & 213s & 112s & 663   \\
\rowcolorA
CVE-2018-17360 & 3645s & 5071s  & 91s  & 980s  & 981s  & 47s & 285s & T.O.\tnote{3} & 101s & T.O.  \\
CVE-2017-7303  & 3943s & 4652s  & 68s & 1092s & 1075s & 66s & 168s & 957s & 87s  & 1278s \\
\rowcolorA
CVE-2017-14940 & 3848s & 4795s  & 67s & 1131s & 1127s & 67s & 175s & 150s & 91s  & 483s  \\
CVE-2017-16828 & 623s  & 141s   & 61s & 113s  & 123s  & 40s & 162s & 72s  & 89s  & 363s  \\ 
\hline
% \midrule
\textbf{Average} & 1647s & 4477s & 81s & 339s & 1577s & 47s & 183s & 280s & 106s & 616s \\
\hline
\bottomrule
\end{tabular}
% \end{adjustbox}

\begin{tablenotes}[para]
\footnotesize
\item[1] \textbf{SelectFuzz: TS} - Temporal-Specialization
\item[2] \textbf{\tool{}: SA} - static analysis for FCC; \textbf{Opt} - seed optimization for reachable seeds;
\textbf{Mutator} - mutator generation for bug-specific mutators.
\item[3] \textbf{T.O.}: Timeout (3600 seconds; it is not counted when calculating the average time). 
\end{tablenotes}
\end{threeparttable}
}

\label{tab:preparation}

\end{table*}

%% file: table/Ablation_Seed.tex
\begin{table*}[!t]
\centering
\caption{Time to Bugs With Different Sets of Initial Seeds. Seeds generated by \tool{} can trigger bugs the most quickly.}

% \begin{adjustbox}{width=\textwidth,center}
\resizebox{0.95\linewidth}{!}{
\begin{threeparttable}

\begin{tabular}{l|ccc|ccc|ccc|ccc}
\toprule
\hline
 \multirow{2}{*}{\textbf{CVE ID}} &
  \multicolumn{3}{c|}{\textbf{AFLGo}} &
  \multicolumn{3}{c|}{\textbf{Beacon}} &
  \multicolumn{3}{c|}{\textbf{WindRanger}} &
  \multicolumn{3}{c}{\textbf{SelectFuzz}} \\ \cline{2-13} 
&
  \textbf{OS}\tnote{1} &
  \textbf{SS}\tnote{2} &
  \textbf{RLS}\tnote{3} &
  \textbf{OS} &
  \textbf{SS} &
  \textbf{RLS} &
  \textbf{OS} &
  \textbf{SS} &
  \textbf{RLS} &
  \textbf{OS} &
  \textbf{SS} &
  \textbf{RLS} \\ \hline
  \rowcolorA
CVE-2018-14498 &
  59161s &
  37525s &
  \textbf{34s} &
  T.O.\tnote{4} &
  T.O. &
  T.O. &
  T.O. &
  6353s &
  \textbf{11s} &
  T.O. &
  12921s &
  \textbf{40s} \\
CVE-2020-13790 &
  81026s &
  84509s &
  \textbf{17828s} &
  T.O. &
  T.O. &
  T.O. &
  T.O. &
  T.O. &
  T.O. &
  T.O. &
  T.O. &
  T.O. \\
  \rowcolorA
CVE-2016-4487 &
  1272s &
  1488s &
  \textbf{22s} &
  276s &
  180s &
  \textbf{3s} &
  308s &
  1497s &
  \textbf{23s} &
  423s &
  316s &
  \textbf{94s} \\
CVE-2016-4489 &
  2717s &
  2420s &
  \textbf{1682s} &
  396s &
  384s &
  \textbf{272s} &
  641s &
  470s &
  \textbf{398s} &
  3465s &
  \textbf{916s} &
  1600s \\
  \rowcolorA
CVE-2016-4491 &
  7514s &
  6466s &
  \textbf{9s} &
  2817s &
  2194s &
  \textbf{5s} &
  12039s &
  12874s &
  \textbf{376s} &
  13323s &
  \textbf{12734s} &
  13114s \\
CVE-2017-8393 &
  2875s &
  1379s &
  \textbf{73s} &
  \textbf{66901s} &
  T.O. &
  T.O. &
  2641s &
  \textbf{354s} &
  435s &
  3184s &
  2100s &
  \textbf{1487s} \\
  \rowcolorA
CVE-2017-8394 &
  2077s &
  12046s &
  \textbf{51s} &
  87s &
  53258s &
  \textbf{75s} &
  2201s &
  78775s &
  \textbf{170s} &
  \textbf{480s} &
  T.O. &
  3847s \\
CVE-2017-8395 &
  230s &
  186s &
  \textbf{14s} &
  26s&
  40s &
  \textbf{4s} &
  163s &
  282s &
  \textbf{16s} &
  309s &
  T.O. &
  \textbf{17s} \\
  \rowcolorA
CVE-2017-8392 &
  574s &
  225s &
  \textbf{27s} &
  79490s &
  \textbf{5428s} &
  33207s &
  380s &
  97s &
  \textbf{18s} &
  313s &
  197s &
  \textbf{11s} \\
CVE-2017-8397 &
  14656s &
  855s &
  \textbf{764s} &
  20739s &
  \textbf{80s} &
  146s &
  27054s &
  367s &
  \textbf{83s} &
  T.O. &
  \textbf{508s} &
  T.O. \\
  \rowcolorA
CVE-2018-17360 &
  \textbf{55014s} &
  T.O. &
  T.O. &
  \textbf{6611s} &
  T.O. &
  T.O. &
  \textbf{9734s} &
  T.O. &
  T.O. &
  \textbf{11168s} &
  T.O. &
  T.O. \\
CVE-2017-7303 &
  \textbf{365s} &
  6408s &
  13320s &
  \textbf{34s} &
  842s &
  2203s &
  5495s &
  4607s &
  \textbf{1614s} &
  \textbf{253s} &
  6116s &
  17273s \\
  \rowcolorA
CVE-2017-14940 &
  T.O. &
  T.O. &
  T.O. &
  T.O. &
  T.O. &
  T.O. &
  T.O. &
  T.O. &
  T.O. &
  T.O. &
  T.O. &
  T.O. \\
CVE-2017-16828 &
  \textbf{547s} &
  694s &
  2419s &
  25055s &
  23221s &
  \textbf{2136s} &
  \textbf{193s} &
  4517s &
  20978s &
  T.O. &
  T.O. &
  \textbf{55485s} \\ 
  \hline
  \textbf{\# of Fastest}\tnote{5} & 3 & 0 & 10 & 3 & 2& 6 & 2 & 1 & 9 & 3 & 3 & 6 \\
  \hline
  \bottomrule
\end{tabular}

% \end{adjustbox}

\begin{tablenotes}[para]
\footnotesize
\item[1] \textbf{OS}: Original Seeds; \item[2] \textbf{SS}: Simple Seeds; \item[3] \textbf{RLS}: Seeds Provided by \tool{}.
\item[4] \textbf{T.O.}: Timeout, indicating that the target vulnerability was not triggered within the specified time (24 hours = 86400 seconds).
\item[5] The number of vulnerabilities the fuzzer triggers most quickly.
\end{tablenotes}
\end{threeparttable}
}

\label{tab:ablation_seed}

\end{table*}

%% file: table/Ablation_Mutation.tex
\begin{table}[!t]
\centering
\caption{Time to Bugs for Different Fuzzers, with \tool-generated seeds.}
% \begin{adjustbox}{width=1\columnwidth,center}
\resizebox{\linewidth}{!}{
\begin{threeparttable}

\begin{tabular}{l|ccccc}
\toprule
\hline
\textbf{CVE ID} & \rotatebox{60}{\textbf{AFLGo}} & \rotatebox{60}{\textbf{Beacon}} & \rotatebox{60}{\textbf{WindRanger}} & \rotatebox{60}{\textbf{SelectFuzz}}  & \rotatebox{60}{\textbf{\tool}}     \\ 
\hline
\rowcolorA
CVE-2018-14498 &
  34s &
  T.O.\tnote{1} &
  \textbf{11s} &
  40s &
  23s \\
CVE-2020-13790 &
  17828s &
  T.O. &
  T.O. &
  T.O. &
  \textbf{6932s} \\
\rowcolorA
CVE-2016-4487 &
  22s &
  \textbf{3s} &
  23s &
  94s &
  5s \\
CVE-2016-4489 &
  1682s &
  272s &
  398s &
  1600s &
  \textbf{268s} \\
  \rowcolorA
CVE-2016-4491 &
  9s &
  5s &
  376s &
  13114s &
  \textbf{2s} \\
CVE-2017-8393 &
  73s &
  T.O. &
  435s &
  1487s &
  \textbf{59s} \\
  \rowcolorA
CVE-2017-8394 &
  51s &
  75s &
  170s &
  3847s &
  \textbf{19s} \\
CVE-2017-8395 &
  14s &
  \textbf{4s} &
  16s &
  17s &
  12s \\
  \rowcolorA
CVE-2017-8392  & 27s            & 33207s          & 18s                 & \textbf{11s} & \textbf{11s} \\
CVE-2017-8397 &
  764s &
  146s &
  83s &
  T.O. &
  \textbf{47s} \\
  \rowcolorA
CVE-2018-17360 &
  T.O. &
  T.O. &
  T.O. &
  T.O. &
  T.O. \\
CVE-2017-7303 &
  13320s &
  2203s &
  \textbf{1614s} &
  17273s &
  12420s \\
  \rowcolorA
CVE-2017-14940 &
  \textbf{26s} &
  T.O. &
  139s &
  T.O. &
  458s \\
CVE-2017-16828 &
  2419s &
  \textbf{2136s} &
  20978s &
  55485s &
  2313s \\
  \hline
  \textbf{\# of Fastest}\tnote{2} & 1 & 3 & 2 & 1 & 7 \\
  \hline
  \bottomrule
\end{tabular}
\begin{tablenotes}
\footnotesize
\item[1] \textbf{T.O.}: Timeout (24 hours = 86400 seconds).
\item[2] The number of vulnerabilities the fuzzer triggers most quickly.
\end{tablenotes}

\end{threeparttable}
}

% \end{adjustbox}
\label{tab:ablation_mutation}
\end{table}

%% file: table/Ablation_LLM_Model.tex
% Please add the following required packages to your document preamble:
% \usepackage{multirow}
\begin{table*}[!t]
\centering
\caption{Influence of Different LLM Models}

\resizebox{\linewidth}{!}{
\begin{threeparttable}
\begin{tabular}{ll|ccc|ccc|ccc|ccc}
\toprule
\hline
\multirow{2}{*}{\textbf{Program}} &
  \multirow{2}{*}{\textbf{CVE ID}} &
  \multicolumn{3}{c|}{\textbf{GPT-4-0125-preview}} &
  \multicolumn{3}{c|}{\textbf{GPT-4o}} &
  \multicolumn{3}{c|}{\textbf{GPT-o1-mini-2024-09-12}} &
  \multicolumn{3}{c}{\textbf{GPT-o1-preview-2024-09-12}} \\ \cline{3-14} 
        &            & RAG\tnote{1}  & Opt\tnote{2}   & Mutator\tnote{3} & RAG  & Opt  & Mutator & RAG  & Opt  & Mutator & RAG   & Opt   & Mutator \\ \hline
\rowcolorA cjpeg   & CVE-2020-13790 & 134s & T.O.\tnote{4}  & 88s      & 64s  & 549s & 61s      & 88s  & 238s & 103s     & 506s  & 2176s & 312s     \\
cxxfilt & CVE-2016-4487  & 210s & 1548s & 122s     & 191s & 102s & 126s     & 142s & 135s & 142s     & 1020s & 793s  & 338s     \\
\rowcolorA objcopy & CVE-2017-8393  & 222s & 562s  & 143s     & 134s & 210s & 122s     & 181s & 412s & 98s      & 719s  & 1908s & 442s     \\
objdump & CVE-2017-8392  & 291s & 1450s & 107s     & 283s & 622s & 94s      & 203s & 903s & 134s     & 1103s & T.O.  & 293s     \\ \hline
\end{tabular}

\begin{tablenotes}[para]
\footnotesize
\item[1] \textbf{RAG}: command option via RAG.
\item[2] \textbf{Opt}: seed optimization for reachable seeds.
\item[3] \textbf{Mutator}: mutator generation for bug-specific mutators.
\item[4] \textbf{T.O.}: timeout (3600 seconds). 
\end{tablenotes}
\end{threeparttable}
}

\label{tab:llm_model}

\end{table*}

%% file: body/discussion.tex
\section{Discussion} \label{sec:discussion}

In this section, we discuss the insights gained from applying \tool{} to real-world vulnerability projects and its interactions with LLMs.
% We present the limitations of current vulnerability report formats, the complexities involved in generating diverse input data for testing, and the reasoning capabilities of LLMs that enable \tool{} to handle vulnerabilities beyond the LLMs' training data.
Each of these factors influences the effectiveness of \tool{} and highlights areas for future improvement in automated vulnerability testing.

\subsection{Description In CVE Report}

Unlike other fuzzing methods requiring manual analysis, \tool{} locates vulnerabilities by analyzing bug reports, reducing the manual workload. However, this requires the bug reports to specify the vulnerability's location at the function-level. For instance, CVE-2016-4487 only mentions the \texttt{btypevec} variable but does not identify the vulnerable function, which is actually \texttt{remember\_Ktype}. As a result, \tool{} cannot locate the vulnerability, hindering further testing. Additionally, bug reports may mention functions that trigger the vulnerability or were patched, not necessarily the vulnerable function itself, leading to incorrect localization and unnecessary tests.

\subsection{Generating Complex Inputs}

In section~\ref{sec:input_generation}, \tool{} can generate simple string inputs via LLMs, but more complex inputs, like images, require Python scripts. We also encountered challenges with tools like \texttt{readelf}, \texttt{objdump}, and \texttt{objcopy}, which require ELF files or compiled programs. Although we asked the LLM to generate Python scripts for creating ELF files or compiling programs, the complexity of these files often caused the scripts to fail. By instructing the LLM to generate Python code that creates a \texttt{.c} file and compiles it locally, we successfully generated inputs requiring compilation. However, this approach struggles with even more complex inputs, like video or 3D modeling files. Future work will focus on optimizing input generation for more complex formats, such as those involving multimodal generative models or multimodal LLMs~\cite{hu2024terrain,lu2024autoregressive,zhang2024mm}. 

\subsection{Reasoning Capability of LLM}

Given that large language models operate based on extensive text data as training sets, we concern that LLMs might rely on their memory rather than the inference capabilities when assisting \tool{}.
To investigate this, we design an experiment using a CVE vulnerability disclosed after the latest update to the model's knowledge base. We select GPT-4o to test the vulnerability CVE-2024-34459. GPT-4o's training dataset was last updated in October 2023, whereas information and patches for CVE-2024-34459 were only released in May 2024.
The vulnerability is a heap-buffer-overflow and exists in the program \texttt{xmlint}, where the vulnerable function is \texttt{xmllint.c:xmlHTMLPrintFileContext()}.

GPT-4o successfully completes the tasks of \textit{SA} (46 seconds), \textit{RAG} (157 seconds), \textit{Opt} (436 seconds), and \textit{Mutator} (113 seconds). 
As a result, \tool{} succeeds in triggering the bug in 941 seconds.
The results show that with GPT-4o, \tool{} can effectively handle vulnerabilities not included in the LLM's training data. Specifically, for CVE-2024-34459, which was disclosed after GPT-4o’s latest update, \tool{} successfully analyzes and optimizes seeds for this newly disclosed vulnerability. This also demonstrates the GPT-4o’s inference ability, rather than solely relying on its memory of historical data, which is crucial in assisting \tool{} to identify vulnerabilities.

%% file: body/relate.tex
\section{Related Work} \label{sec:relate}

% \todo{to evaluate technique rather than implementation.}

In this paper, we utilize LLMs to assist directed fuzzing, by using the capability of LLMs to understand code and bugs. 
The existing research on directed fuzzing can be classified into three categories, including optimization of the guidance towards target locations~\cite{bohme2017directed, huang2024titan, xiang2024critical}, reduction of unnecessary exploration of unrelated code~\cite{huang2024everything, li2024sdfuzz, luo2023selectfuzz}, and directed fuzzing in specific scenarios~\cite{tan2023syzdirect, yuan2023ddrace, cao2023oddfuzz}.
Another related research is to use LLMs in assisting fuzzing, and the existing fuzzers focus on understanding the formats of inputs~\cite{deng2023large, wang2024llmif, deng2024large}.

\noindent\textbf{Optimization of Guidance for Directed Fuzzing.}
Since the first directed fuzzer AFLGo~\cite{bohme2017directed}, researchers have developed various methods to optimize the guidance towards target locations. Hawkeye~\cite{chen2018hawkeye} improves AFLGo by considering multiple paths that can reach target locations. CAFL~\cite{lee2021constraint} then realizes that constraints in programs also impact the efficiency of reaching target locations. $MC^2$~\cite{shah2022mc2} introduces a performance metric, the number of oracle queries, to guide directed fuzzing. 
Another fuzzer 1dFuzz~\cite{yang20231dfuzz} also introduces an extra metric, the trailing call sequence, to guide directed fuzzing to follow specific function call sequences.
Similarly, PDGF~\cite{zhang2024predecessor} introduces a new metric, the regional maturity, to improve efficiency by gradually reaching more predecessors.
WindRanger~\cite{du2022windranger} identifies basic blocks that deviate from target locations so that fuzzing can focus more on execution paths that reach target locations. 
Titan~\cite{huang2024titan} and WAFLGO~\cite{xiang2024critical} improves the efficiency of directed fuzzing via regarding it as a multi-objective optimization problem.
DeepGo~\cite{lin2024deepgo} improves fuzzing efficiency via using reinforcement learning.
These fuzzers utilize the knowledge from experts and improve fuzzing efficiency based on different understandings of programs.
However, they overlook the manual efforts required to successfully reach target locations.

\noindent\textbf{Reduction of Unrelated Exploration for Directed Fuzzing.}
While the optimization of fuzzing process will prefer seeds that are more likely to reach target locations, it still examines a large number of inputs that explore unrelated code regions. Therefore, researchers either trim the size of generated inputs or use selective instrumentation to improve efficiency~\cite{zong2020fuzzguard, huang2024everything, luo2023selectfuzz}.
To trim the size of generated inputs, FuzzGuard~\cite{zong2020fuzzguard} uses deep learning to predict if an input can reach target locations. Another fuzzer Halo~\cite{huang2024everything} restrains input generation based on likely invariants so that more reachable inputs are generated. 
Selective instrumentation only instruments related code regions so that fuzzing will only be guided based on related code~\cite{luo2023selectfuzz, li2024sdfuzz}.
Another solution is to terminate the execution of a program if it reaches unrelated code regions~\cite{huang2022beacon}.
These fuzzers require accurate analysis of target programs so that the reduction of unrelated exploration does not miss related code regions.
However, the accuracy may lead to heavy analysis of target programs.
Moreover, they use pre-defined mutation operators, which do not connect to target bugs, impairing the efficiency of bug detection.

\noindent\textbf{Directed Fuzzing in Specific Scenarios.}
Directed fuzzing is also used in different applications.
To fuzz Linux kernel, SyzDirect~\cite{tan2023syzdirect} uses scalable static analysis to identify vulnerable information, which is used to guide the directed fuzzing. 
To detect concurrency use-after-free vulnerabilities, DDRace~\cite{yuan2023ddrace} identifies potential use-after-free locations as target locations and guides fuzzing to gradually satisfy those target sites.
To detect Java deserialization vulnerabilities, ODDFuzz~\cite{cao2023oddfuzz} identifies
candidate gadget chains and guides fuzzing to examine each target site in the gadget chains.
To detect vulnerabilities in Internet of Things (IoT) firmware, LABRADOR~\cite{liu2024labrador} deduces code coverage in firmware and estimates the distance to target sensitive code.
These fuzzers always require deep understanding of target programs or vulnerabilities. However, such understanding is usually obtained from experts. For example, the patterns of target sites for use-after-free and deserialization vulnerabilities are identified based on expertise in those specific vulnerabilities.

\noindent\textbf{LLM for Fuzzing.}
Existing research on using LLMs for fuzzing concentrates on generating appropriate inputs for target programs.
They use the excellent capability of LLMs to understand the specifications for inputs or learn input format during their training.
TitanFuzz~\cite{deng2023large} uses the LLMs' capability of code generation and generates programs to invoke APIs in deep learning libraries.
Some fuzzers extract protocol information from the specification so that fuzzers can generate valid and various messages to test protocol code~\cite{wang2024llmif, ma2024one}.
For well-known protocols, the format of input message can be obtained by directly query LLMs~\cite{meng2024large}.
Similarly, LLMs can also generate effective inputs for programs written in widely-used programming languages~\cite{xia2024fuzz4all, kang2023large}.
Some other fuzzers explore the capability of LLMs to generate edge or unusual inputs for target programs~\cite{deng2024large, Zhang2025your}.
However, these fuzzers focus on the format of inputs, ignoring the potential of LLMs in other stages of fuzzing.

%% file: body/conclusion.tex
\section{Conclusion} \label{sec:conclusion}

The dilemma of randomness in fuzzing motivates us to reduce the randomness while maintaining the effectiveness of fuzzing.
We identify two key components, initial seeds and mutators, that are significantly impacted by randomness and propose solutions to solve the problems, respectively.
LLM is crucial to the success of our \tool{}.
We use LLM to remove the randomness in initial seeds, generating reachable seeds.
LLM is also used to reduce the randomness in mutators, constructing bug-specific mutators.
The experiment results show that both the reachable seeds and bug-specific mutators are efficient in exposing bugs.